\documentclass[structabstract]{aa}  
\usepackage{multirow,bigdelim}
\usepackage{graphicx,subfigure}
\usepackage{lscape}
\usepackage{longtable}
\usepackage{natbib}
\bibpunct{(}{)}{;}{a}{}{,}
\usepackage{txfonts}
\usepackage{caption}

\begin{document}
%
 \title{A $\lambda=3$\,mm molecular line survey of NGC\,1068.\\ Chemical signatures of an AGN environment}



  \author{R. Aladro
          \inst{1,2},
S. Viti\inst{2}, E. Bayet\inst{3}, D. Riquelme\inst{4}, S. Mart\'in\inst{1}, R. Mauersberger\inst{5}, J. Mart\'in-Pintado\inst{6}, M. A. Requena-Torres\inst{4}, C. Kramer\inst{7}, A. Wei{\ss}\inst{4}  }

   \institute{European Southern Observatory, Avda. Alonso de C\'ordova 3107, Vitacura, Santiago, Chile.
        \email{raladro@eso.org}
 	\and
 	University College London (UCL),
              Dept. of Physics \&  Astronomy, Gower Street, London WC1E 6BT, UK.
 	\and
	Sub-Department of Astrophysics, University of Oxford, Denys Wilkinson Building, Keble Road, Oxford OX1 3RH, UK.
         \and
	Max-Planck-Institut f\"{u}r Radioastronomie, Auf dem H\"{u}gel 69, 53121 Bonn, Germany
	\and
	Joint ALMA Observatory,  Avda. Alonso de C\'ordova 3107, Vitacura, Santiago, Chile.
        \and
	Centro de Astrobiolog\'ia (CSIC-INTA), Ctra. de Torrej\'on Ajalvir km 4, E-28850 Torrej\'on de Ardoz, Madrid, Spain.
	\and
	Instituto de Radioastronom\'ia Milim\'etrica, Avda. Divina Pastora, 7, Local 20, E-18012 Granada, Spain.
}

  \date{Received  /  Accepted }


 \abstract
{}
{To study the molecular composition of the interstellar medium (ISM) surrounding an Active Galactic Nucleus (AGN), by making an inventory of molecular species and their abundances. To establish a chemical differentiation between starburst galaxies and AGN.}
{We used the IRAM-30\,m telescope to observe the central 1.5-2\,kpc region of NGC\,1068, covering the frequencies between 86.2\,GHz and 115.6\,GHz. Using Boltzmann diagrams, we calculated the column densities of the detected molecules. We used a chemical model to reproduce the abundances found in the AGN, to determine the origin of each detected species, and to test the influence of UV fields, cosmic rays, and shocks on the ISM.} 
{We identified 24 different molecular species and isotopologues, among which HC$_3$N, SO, N$_2$H$^+$, CH$_3$CN, NS, $^{13}$CN, and HN$^{13}$C are detected for the first time in NGC\,1068. A comparison of the abundances in the nuclear regions of NGC\,1068, M\,82 and NGC\,253 allowed us to establish a chemical differentiation between starburst galaxies and AGN. H$_2$CO and CH$_3$CCH, two abundant species in starburst galaxies, are not detected in NGC\,1068, probably because they are destroyed by UV fields or shocks. On the other hand, species such as CN, SiO, HCO$^+$ and HCN, are enhanced by cosmic ray radiation fields. We obtained the upper limits to the isotopic ratios $^{12}$C\,/\,$^{13}$C=49, $^{16}$O\,/\,$^{18}$O=177 and  $^{32}$S\,/\,$^{34}$S=5. These ratios are much lower in this AGN than in starburst galaxies. Our chemical models suggest that the chemistry in the nucleus of NGC\,1068 is strongly influenced by cosmic rays, although high values of both cosmic rays and far ultraviolet (FUV) radiation 
fields also explain well the observations. C-shocks can explain the abundances of C$_2$H and H$_2$CO, but do not strongly affect the abundances of the other detected species.} 
{The gas in the nucleus of NGC\,1068 has a  different chemical composition as compared to starburst galaxies. The distinct physical processes dominating galaxy nuclei (e.g. C-shocks, UV fields, X-rays, cosmic rays) leave clear imprints in the chemistry of the gas, which allow to characterise the nucleus activity by its molecular abundances.}

\authorrunning{Aladro et al. (2012)}
\titlerunning{A $\lambda=3$\,mm molecular line survey of NGC\,1068}
  \maketitle{}


\section{Introduction}
\label{sect.Intro}

\begin{figure*}[t!]
        \centering
        \subfigure[]{\label{12COmap}\includegraphics[width=60mm]{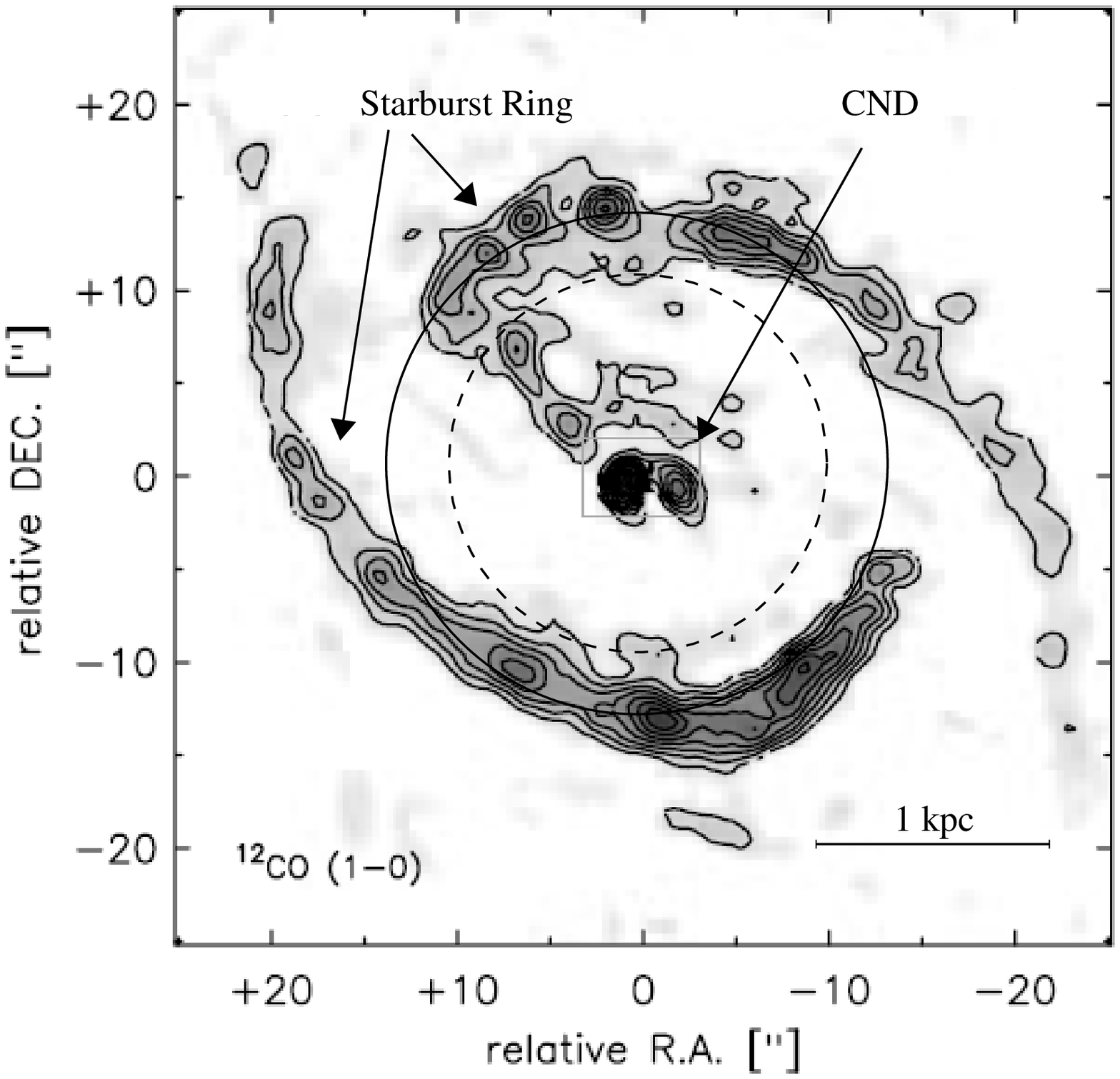}}
                %
        \subfigure[]{\label{3mmsurvey}\includegraphics[width=102mm]{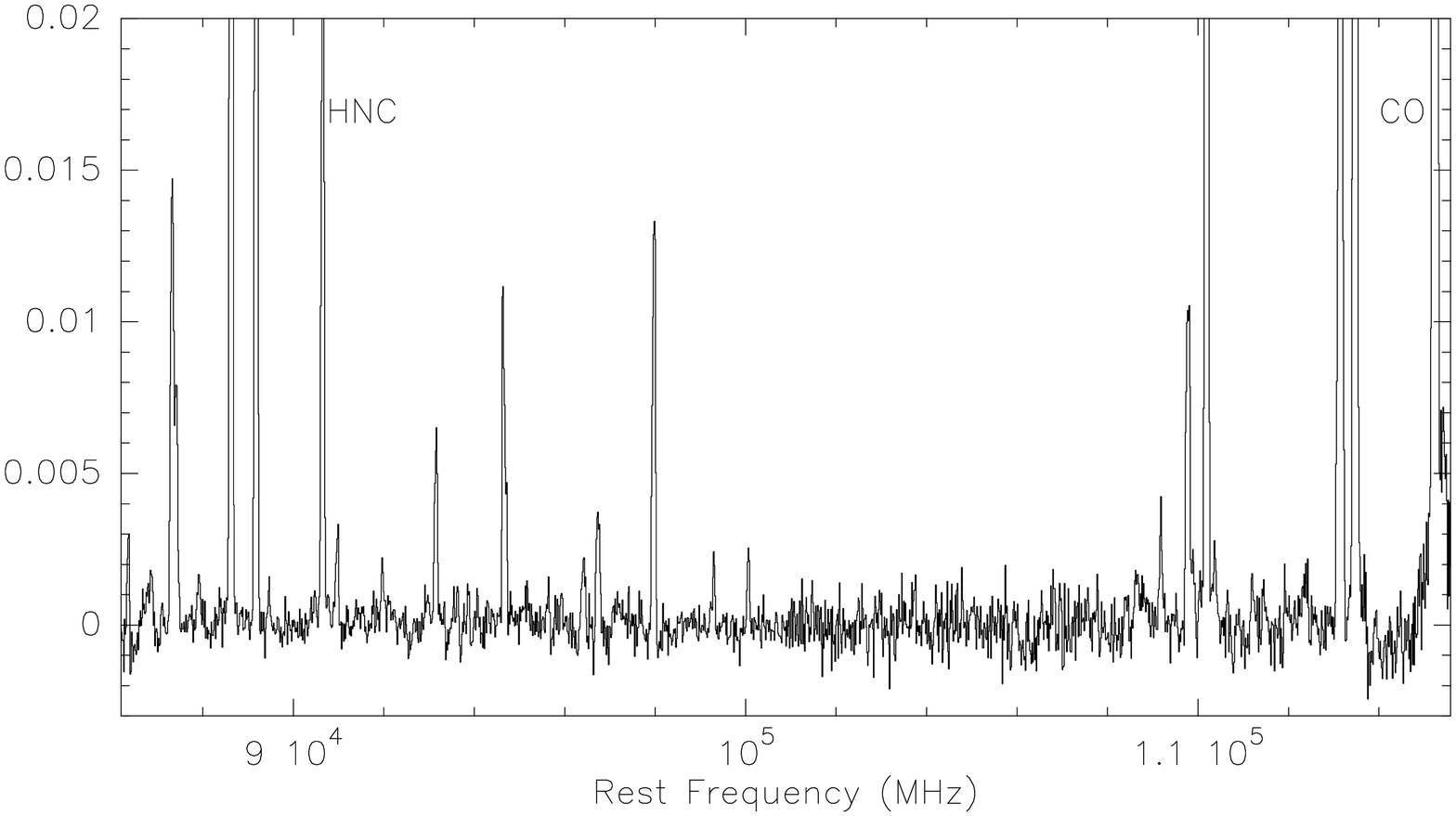}}
        \caption{(a) $^{12}$CO$\,(1-0)$ map of the NGC\,1068 nucleus (taken from \citealt{Schinnerer00}). The beam sizes at the lowest and highest frequencies of the survey are plotted with a solid and dotted circles respectively. The starburst ring and CND are also indicated. (b) Our molecular line survey in the 3\,mm wavelengths. Temperatures are in $T_{\rm MB}$(K).
}\label{12CO}
\end{figure*}

NGC\,1068 is one of the closest (D=14.4\,Mpc, 1$''=72$\,pc, \citealt{Bland-Hawthorn97}) Seyfert 2 galaxies. It has an infrared (IR) luminosity of $3\times10^{11}$L$_\odot$ \citep{Telesco80}. IR observations reveal warm dust with temperatures of 320\,K surrounding a smaller hot structure \citep{Jaffe04} assumed to be a central black hole of mass $1.7\times 10^7$M$_\odot$ \citep{Schinnerer00}. CO observations show molecular spiral arms forming a starburst ring 10$''$ from the nucleus. Inside the ring, NGC\,1068 has a circumnuclear (CND) disk with a thickness of 10\,pc (\citealt{Schinnerer00}, see Fig.~\ref{12CO}).

NGC\,1068 is therefore one of the best extragalactic targets to study the physical and chemical properties of the interstellar medium (ISM) in the vicinity of an active galactic nucleus (AGN). 
The molecular material close to the central engines of AGN is pervaded by X-ray and cosmic ray radiations originating in the nuclear accretion disks \citep{Maloney96}. Previous single-dish and interferometric observations of NGC\,1068 in a number of molecules, such as CO, SiO, HOC$^+$ and CN, have revealed a chemistry interpreted as the result of a giant X-ray dominated region (XDR) in its nucleus, where shocks could also be heating the gas \citep{Usero04,Burillo10,Kamenetzky11}. This strong X-ray radiation has also been claimed to be responsible for the enhanced HCN-to-HCO$^+$ line ratio, which differs between AGN and starburst environments \citep{Kohno01,Krips08}. 
However, a comprehensive study of the molecular content of the ISM close to AGNs has not been carried out so far. As a consequence, the influence of X-rays and cosmic rays on the molecular gas is only partially understood.

Unbiased molecular line surveys done toward the nuclei of the starburst galaxies NGC\,253 \citep{Martin06b} and M\,82 \citep{Aladro11} have allowed to determine the impact of large-scale shocks and UV fields on the ISM, as well as to analyse how its composition changes as the starburst phenomena evolve with time.
It is clear from these studies that a detailed molecular inventory description of a prototypical AGN is also needed in order to understand the properties of extragalactic XDRs, and also to chemically differentiate between nuclear powering sources (starbursts versus AGN). 

In this work we present a comprehensive chemical study of the gas in the NGC\,1068 nucleus. We characterise its physical and chemical properties, and highlight some important differences with the massive star-forming regions in starburst galaxies. In Sects.~\ref{sect.Obs} and ~\ref{analysis} we present details of the observations, data reduction and analysis. Sect.~\ref{sect.results} shows the results obtained from the survey, and a discussion on the NGC\,1068 chemistry. In Sect.~\ref{modelling} we present a chemical model of NGC\,1068 used to investigate how UV radiation, cosmic rays and C-type shocks may affect its gas. We also discuss the origin of each detected species and present the comparison of the model results with the observational data. In Sect.~\ref{comparison} we compare the properties of the circumnuclear gas in NGC\,1068 with those found in starburst galaxies. Finally, we present our conclusions in Sect.~\ref{conclusions}.

\section{Observations and data reduction}
\label{sect.Obs}

The observations were carried out with the IRAM-30\,m telescope\footnote{IRAM is supported by INSU/CNRS (France), MPG (Germany) and IGN (Spain).} (Pico Veleta Observatory, Spain) between October 2009 and July 2010. We observed the nucleus of NGC1068, at a nominal position $\alpha_{2000}=$\,02:42:40.9, $\delta_{2000}$=\,-00:00:46.0, between the frequencies 86.2 and 115.6\,GHz. The assumed heliocentric systemic velocity was 1110\,km\,s$^{-1}$. The half-power beam width (HPBW) ranged from 29$''$ to 21$''$, corresponding to a spatial scale of 2-1.5\,kpc. We used the band E0 of the EMIR receiver and the WILMA autocorrelator. This receiver-backend configuration allowed us to cover 8\,GHz simultaneously in the vertical and horizontal polarisations, and led to a channel width spacing of $7-9$\,km\,s$^{-1}$. The IRAM-30m data were first calibrated to the antenna temperature ($T_{\rm A}^*$) scale using the chopper-wheel method \citep{Penzias73}. The observations were done wobbling the secondary mirror with a switching 
frequency of 0.5\,Hz 
and a beam throw of 110$''$ in azimuth. We checked the pointing accuracy every hour towards the nearby bright continuum sources 2251+158 and 0113-118. The pointing corrections were always better than 4$''$. The focus was also checked at the beginning of each run and during sunsets.

\begin{figure*}[h!]
\begin{center}

\includegraphics[angle=0,width=0.6\textwidth]{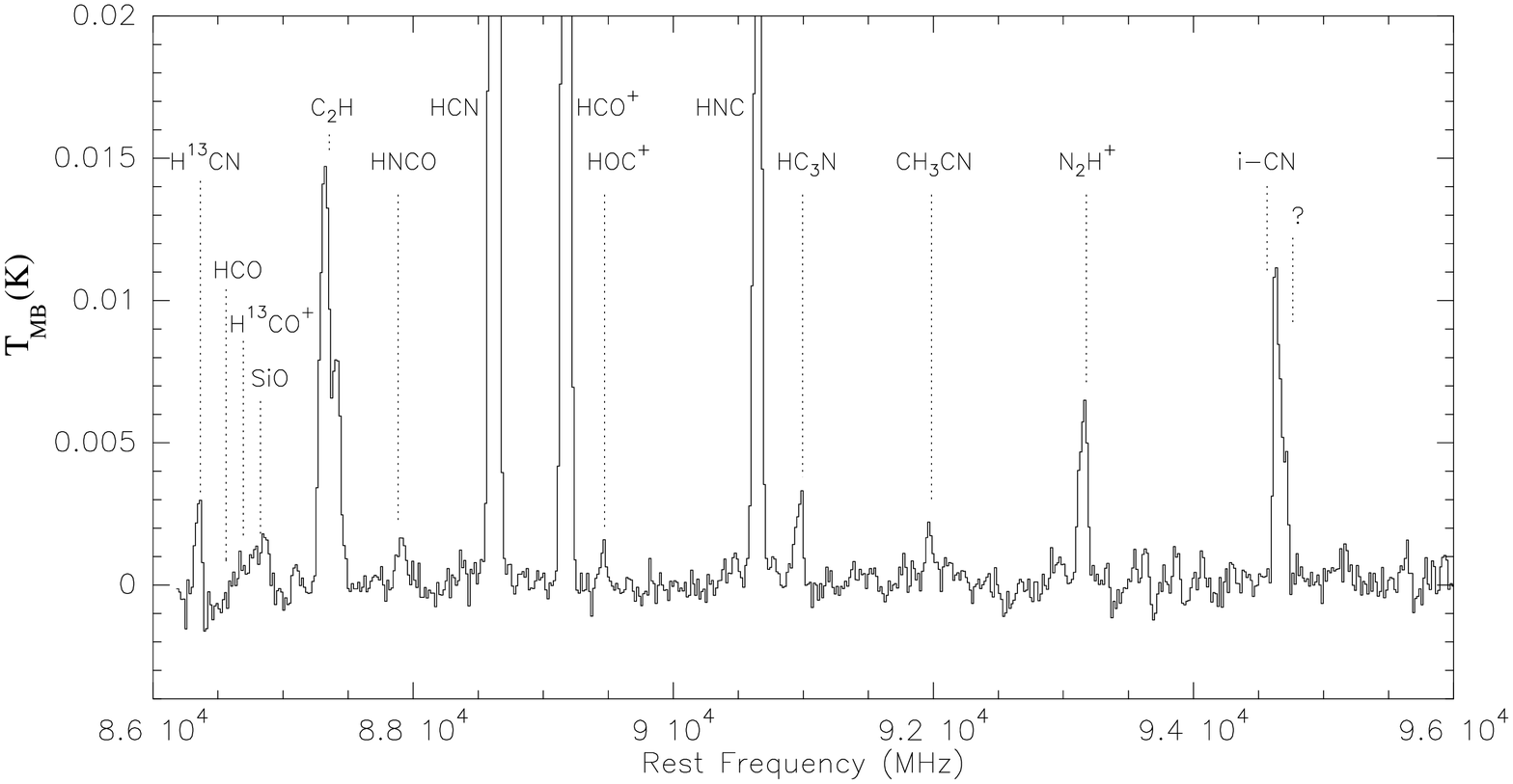}
\includegraphics[angle=0,width=0.6\textwidth]{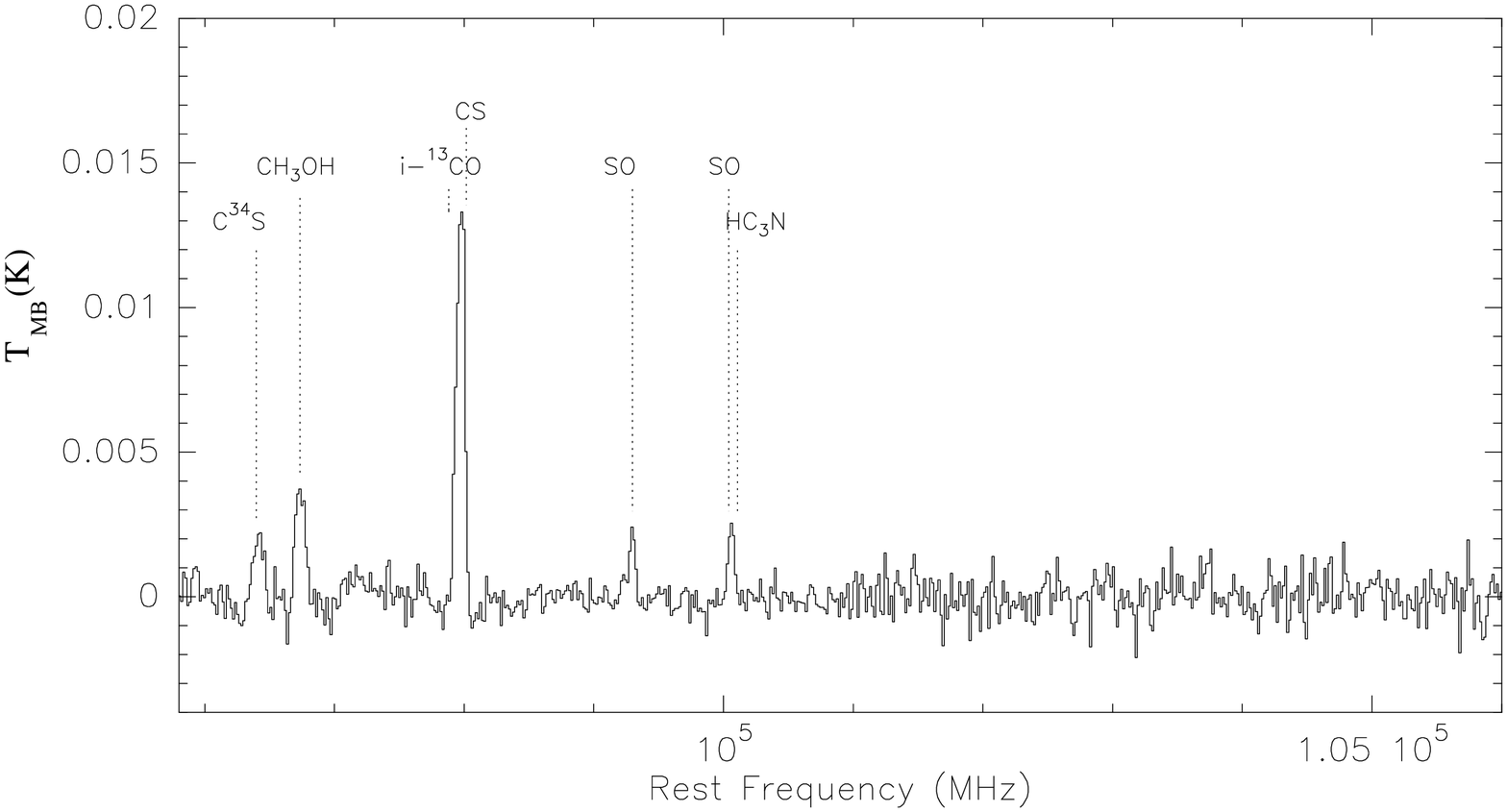}
\includegraphics[angle=0,width=0.6\textwidth]{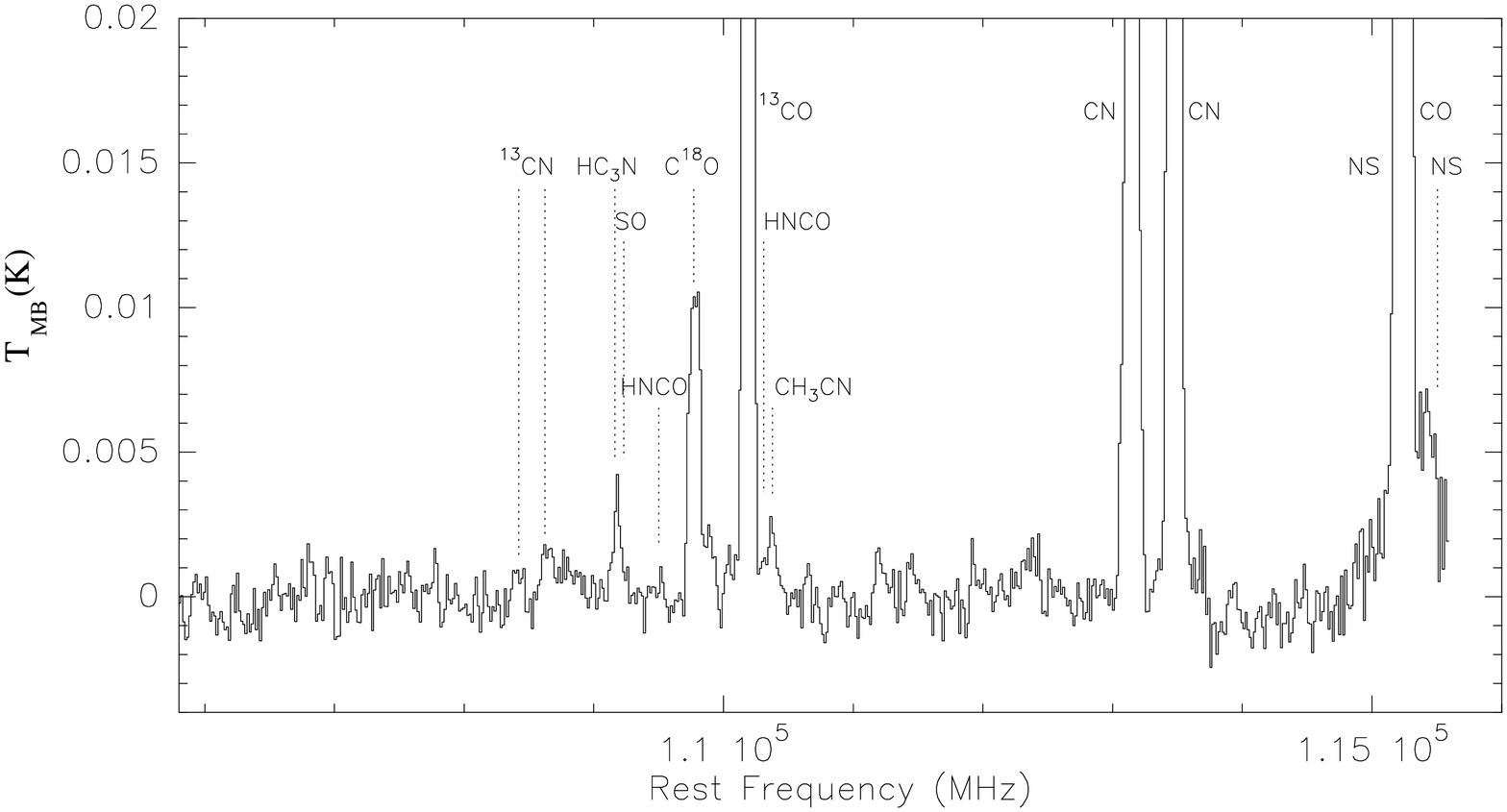}
\caption{NGC\,1068 frequency survey coverage between 86.2 and 115.6\,GHz. The spectra have been smoothed to a velocity resolution of $\sim$44\,km\,s$^{-1}$ (corresponding to $\sim16$\,MHz). The assumed systemic velocity is 1110\,km\,s$^{-1}$. }
\label{Survey123}
\end{center}
\end{figure*}

The observed spectra were converted from $T_{\rm A}^*$ to main beam temperatures ($T_{\rm MB}$) using the relation $T_{\rm MB}=(F_{\rm eff}/B_{\rm eff})\,T_{\rm A}^*$, where $F_{\rm eff}$ is the forward efficiency of the telescope, whose values were between 0.94 and 0.95, and $B_{\rm eff}$ is the main beam efficiency, ranging from 0.77 to 0.81. Linear baselines were subtracted in all cases. We note that some of the spectra still contain faint ripples that were not completely removed by our switching scheme. This results in some of the weak features observed (e.g. around 94\,GHz, 111.3\,GHz and 112.3\,GHz). However, these features are quite below the noise of the spectra. The rms achieved is $\le 2$\,mK ($\le 10$\,mJy/beam) across the whole survey. The survey coverage is shown in Fig.~\ref{3mmsurvey}, while more detailed spectra, where the detected species are labeled, are shown in Fig.~\ref{Survey123}.

The data were also corrected to first order for beam dilution effects as $T_{\rm B}=[(\theta^2_{\rm s}\,+\,\theta^2_{\,\rm b})\,/\,\theta^2_{\,\rm s}]\,T_{\rm MB}$, where $T_{\rm B}$ is the source averaged brightness temperature, $\theta_{\,\rm s}$ is the source size and $\theta_{\,\rm b}$ is the beam size. Based on NGC\,1068 interferometric observations of $^{12}$CO, HCN and $^{13}$CO (e.g. \citealt{Helfer95,Schinnerer00}) we assumed an average source size of 4$''$ for all the species detected in this survey. This value is consistent with other studies focused in the CND of NGC\,1068 (e.g. \citealt{Bayet09b,Krips08,Krips11, Kamenetzky11}).

Gaussian profiles were fitted to all the detected lines. The parameters resulting from those fits are given in Table~\ref{TableGauss}. The reduction of the spectra and Gaussian profile fitting were done using the CLASS \footnote{CLASS $http://www.iram.fr/IRAMFR/GILDAS$} and MASSA\footnote{MASSA $http://damir.iem.csic.es/mediawiki-1.12.0/index.php/MASSA\_User's\_Manual$} software packages. In those cases where there were two or more molecules blended, we performed a synthetic fitting (see \citealt{Martin10,Aladro11} for details about this method) in order to calculate the contribution of each line to the total profile. Each case is further commented in Appendix~\ref{sect.molecules}.

\section{Analysis of the data}
\label{analysis}

The line identification was made using the CDMS \citep{Muller01,Muller05}, JPL \citep{Pickett88} and LOVAS \citep{Lovas92} catalogues, and largely follows the method explained in detail in \citet{Aladro11} for a similar line survey done toward the nucleus of M\,82. The image band rejection of the EMIR receiver was better than 10\,dB along the band, so only $^{13}$CO\,$(1-0)$ and CN\,$(1-0)$ entered from the image band, at 97.9\,GHz and 94.6\,GHz respectively. The latter is blended with one unidentified feature. No strong line is expected at $\sim94.6$\,GHz, and the CN$(1-0)$ line coming from the image band cannot account for the total emission. 
On the other hand, the $^{13}$CO\,$(1-0)$ line coming from the image band is blended with CS\,$(2-1)$. Since there are no other CS lines in our survey, and the exact rejection of the image band at this particular frequency is not known, we cannot calculate the intensity of the CS transition. Unfortunately, it was not observed previously, so we cannot take it from the literature. However, other CS transitions have been observed towards the nucleus of NGC\,1068 by \citet{Bayet09b} and we include them in our study.

\subsection{Contribution of the starburst ring}
\label{SB}

As shown in Fig.~\ref{12COmap},  the angular resolution of the IRAM-30\,m telescope at 3\,mm wavelengths is not high enough to separate the CND of NGC\,1068 from the surrounding starburst ring. As a consequence, some of the detected molecules in our survey might include some emission of the starburst ring picked up the telescope beam. 

The $J-J'=1-0$ transitions of HCO$^+$, HNC, C$^{18}$O, $^{13}$CO and $^{12}$CO are deviated from a Gaussian profile. Instead, they show emission in the wings, and the lines are better fitted by three Gaussian components. As an example, we show in Fig.~\ref{CO} the profile of the CO\,$(1-0)$ line. The gas dynamics in the centre of NGC\,1068, and in particular in the CND, are quite complex. Interferometric observations have already shown that the profiles of CO, HCN and HCO$^+$ lines need dual, triple or even quadrupole Gaussian fits in order to better reproduce their profiles \citep{Krips11}. This was associated to non-circular motions in the central $\sim$100\,pc, where part of the gas might be blown outward due to shocks \citep{Krips11}, while the rest of the gas is accreted toward the AGN along the jet (e.g. \citealt{MullerSanchez09}).

It is not possible, with single-dish observations, to disentangle whether the deviations from the Gaussian profiles are due only to the complex gas dynamics or if the starburst ring emission is also contributing to the line shapes. Furthermore, for other species no cited above there are other factors which make even more difficult to evaluate the possible starburst ring contribution. For example, some molecules such as C$_2$H and HCN have hyperfine structure that intrinsically create deviations from Gaussian profiles. Others, such as H$^{13}$CO$^+$, HC$_3$N or SiO, are blended in our survey or are very weak in NGC\,1068, so a multiple Gaussian fit is not feasible. Therefore, since we cannot correct for the starburst emission, the data we show in the following sections include all the radiation picked up by the telescope beam (i.e. CND plus starburst ring).

\subsection{LTE analysis}
\label{Boltzmann}
Assuming local thermodynamic equilibrium (LTE) and optically thin emission for all the detected species, we made Boltzmann diagrams to estimate the rotational temperatures ($T\rm_{rot}$) and  column densities of the molecules ($N_{\rm mol}$). A detailed explanation of the method can be found in \citet{Goldsmith99}. The rotation diagrams are shown in Fig.~\ref{DR}, while the resulting parameters are listed in Table~\ref{TableNT}. To  determine the rotational temperatures from the Boltzmann plots, at least two transitions of a given molecule are needed. In those cases where only one transition of a molecule lies in our 3\,mm survey (CO, $^{13}$CO, C$^{18}$O, CS, CN, HCO$^+$, HNC and C$^{34}$S), we complemented our observations with other lines, if available, in the literature (see Table~\ref{TableNT} for references).  For  $^{13}$CN, H$^{13}$CN, H$^{13}$CO$^+$ and HN$^{13}$C there are no previous data published, so we simply assumed that these species arise from gas with the same temperature as the main $^{12}
$C isotopologues. We also considered the same temperature for HOC$^+$ and HCO$^+$. For the rest of the cases (C$_2$H, HCO, NS, N$_2$H$^+$) with neither more than one transition in our survey nor in the literature, we fixed a $T_{\rm rot}=10\pm5$\,K, which is an average value of the rotational temperatures determined in this survey.

As mentioned before, the rotation diagrams were done assuming optically thin emission of the gas. However, some molecules are certainly affected by optical thickness (e.g. CO, HCN, \citealt{Jackson93,Papadopoulos99}). Therefore, in these cases we are underestimating the actual column densities. The optical depths of the most intense molecules in the survey are further discussed in Sect.~\ref{compratios}

\section{Results}
\label{sect.results}

\subsection{Detected and undetected molecules}
We detected a total of 35 lines, or groups of lines, in our survey, which correspond to 17 different molecular species and 7  isotopologues. Some of them were detected for the first time in this galaxy, namely HC$_3$N, SO, N$_2$H$^+$,  CH$_3$CN, NS, $^{13}$CN, and HN$^{13}$C. As detailed in Sect.~\ref{sect.Obs}, we complemented our observations with other detections taken from the literature (see Table~\ref{TableNT} for references). From the Boltzmann diagram results, $^{12}$CO, $^{13}$CO, CN and CS show two temperature components, the warmer component coming from the higher $J-J'$ transitions compiled from the literature (marked in Fig.~\ref{DR} with an asterisk). 

We calculated column density ratios that are claimed to be representative of AGNs, as opposite to starburst environments.  The NGC\,1068 chemistry can be summarised as follows: high $N_{\rm HCN}/N_{\rm CO}$  ($\sim3\times10^{-4}$), high $N_{\rm CN}/N_{\rm HCN}$ ($\sim3-4$), low $N_{\rm HCO^+}/N_{\rm HCN}$ ($\sim0.4$), and low $N_{\rm HCO^+}/N_{\rm HOC^+}$ ($\sim48$). These results are consistent with many previous studies, such as \citet{Usero04,Perez09,Krips08,Krips11}. A comparison of these and other molecular ratios with those typically found in starburst galaxies is presented in Sect.~\ref{comparison}.

We also list in  Table~\ref{TableUpper} molecules that were not detected in our survey, but that were successfully observed in other extragalactic sources. We show the 3$\sigma$ upper limits to their column densities, which were derived from the rms noise level of the corresponding spectra. Some of the non-detected species in NGC\,1068, such as methyl acetylene (CH$_3$CCH) or formaldehyde (H$_2$CO), are quite abundant in starburst galaxies.
These two molecules are presumably too faint to be detected in NGC\,1068, as other studies, also covering large ranges of frequencies, did not detect them either (e.g. \citealt{Kamenetzky11}). We discuss in Sect.~\ref{comp}  possible reasons for their weakness.

\subsection{Rotational temperatures}
\label{temperatures}
The rotational temperatures obtained from the Boltzmann diagrams (for those species with more than one detected transition) range between $\sim$3 and $\sim$51\,K (see Table~\ref{TableNT}). These values are quite below the kinetic gas temperature ($T_{\rm kin}$) traced by ammonia (NH$_3$), of $80\pm20$\,K, for the bulk of the molecular gas \citep{Ao2011}. However, we note that $T_{\rm rot}$ are always lower limits to $T_{\rm kin}$, and that the conversion from rotational to kinetic temperatures is not simple.
Also, the data  we use seem to be subthermally excited. 

Our results agree well with previous studies where it was found that low-$J$ CO, HCN, and HCO$^+$ lines trace cold gas ($T_{\rm kin}^{\rm HCO^+}\leq20$\,K, \citealt{Perez09}; $T_{\rm kin}^{\rm CO}\approx30$\,K, \citealt{Israel09}; $T_{\rm kin}^{\rm CO, HCN}\approx50$\,K, \citealt{Sternberg94}).

\begin{table}
\caption{Column densities and rotational temperatures of the observed molecules}
\centering

\begin{tabular}[!h]{lcccc} 
\hline
\hline
Molecule	&	$T_{\rm rot}\dagger$	& 	$N_{\rm mol}$			& Ref 	\\		
		&	K		& 	cm$^{-2}$			& 	 	\\
\hline
$^{12}$CO	&	$5.0\pm0.3$	&	$(3.8\pm0.3)\times10^{18}$ 	& a  		\\
		&	$38.4\pm6.5$	&	$(1.2\pm0.3)\times10^{18}$ 	& a	\\	
$^{13}$CO	&	$4.3\pm0.2$	&	$(3.2\pm0.3)\times10^{17}$ 	& a		\\
		&	$24.1\pm6.3$	&	$(7.8\pm2.8)\times10^{16}$ 	& a		\\
C$^{18}$O	&	$3.3\pm0.1$	&	$(1.0\pm0.1)\times10^{17}$ 	& b 		\\
C$_2$H		&	$10.0\pm5.0$	&	$(1.0\pm0.5)\times10^{16}$ 	& 		\\
CN		&	$8.8\pm0.1$	&	$(5.0\pm0.1)\times10^{15}$ 	& d	 \\
		&	$51.4\pm36.0$	&	$(5.8\pm4.4)\times10^{15}$ 	& d\\
CH$_3$OH	&	$5.6\pm0.1$	&	$(2.3\pm0.1)\times10^{15}$	& 	\\
HCN		&	$5.0\pm0.3$	&	$(1.4\pm0.2)\times10^{15}$	& e	\\
CS		&	$7.1\pm0.4$	&	$(7.6\pm2.2)\times10^{14}$	& c	\\ 
		&	$33.0\pm13.1$	&	$(7.0\pm4.2)\times10^{13}$	& c	\\ 
{\bf{HC$_3$N}}	&	$7.3\pm1.4$	&	$(6.0\pm4.0)\times10^{14}$	& 	 \\	
HCO$^+$		&	$4.9\pm0.1$	&	$(5.3\pm0.2)\times10^{14}$	& e	\\ 
HNC		&	$7.6\pm0.2$	&	$(4.9\pm0.5)\times10^{14}$	& d	\\ 
HCO		& 	$10.0\pm5.0$	& 	$(2.1\pm1.3)\times10^{14}$	& 	\\
{\bf{SO	}}	&	$22.8\pm18.6$	&	$(3.2\pm3.3)\times10^{14}$   	& 	\\ 
{\bf{NS}}		&	$10.0\pm5.0$	&	$(2.8\pm1.4)\times10^{14}$   	& 	\\
HNCO		&	$29.7\pm0.1$	&	$(2.8\pm0.1)\times10^{14}$	& 	\\
{\bf{$^{13}$CN}}	&	$8.8\pm0.1$	&	$(2.2\pm0.1)\times10^{14}$   	& 	 \\
{\bf{N$_2$H$^+$}}	&	$10.0\pm5.0$	&	$(1.1\pm0.5)\times10^{14}$   	& 	\\
C$^{34}$S	&	$3.6\pm0.5$	&	$(1.6\pm0.6)\times10^{14}$   	& b	\\
H$^{13}$CN	&	$5.0\pm0.3$	&	$(8.7\pm0.5)\times10^{13}$	& 	\\
SiO		&	$3.8\pm0.2$	&	$(6.4\pm0.8)\times10^{13}$	& x	\\
{\bf{CH$_3$CN}}	&	$10.2\pm0.1$	&	$(3.9\pm0.1)\times10^{13}$	& 	\\
H$^{13}$CO$^+$	&	$4.9\pm0.1$	&	$(3.1\pm0.1)\times10^{13}$	& 	\\
{\bf{HN$^{13}$C}}	&	$7.6\pm0.2$	&	$(2.1\pm3.2)\times10^{13}$	& 	\\
HOC+		&	$4.9\pm0.1$	&	$(1.1\pm0.6)\times10^{13}$	&	\\
\hline
c$-$C$_3$H$_2$		&	$7.6$		&	$(1.7\pm...)\times10^{13}$		& f	 \\
NH$_3$		&			&	$(1.1\pm0.2)\times10^{14}$	&g \\
\hline
\end{tabular}
\label{TableNT}
\begin{list}{}{}
\item[]Molecules in boldface are new detections in NGC\,1068. The values given in this table are already corrected by beam dilution effects.
\item[]$\dagger$ For those molecules with only one rotational transition available, we assumed $T_{\rm rot}=10\pm5$\,K. See Sect.~\ref{sect.Obs} for details.
\item[] Last column refers to detections taken from the literature: $^a$\,\citet{Israel09},$^b$\,  \citet{Martin09a}, 
$^c$\, \citet{Bayet09b}, $^d$\,\citet{Perez09}, $^e$\,\citet{Krips08}, $^x$\,\citet{Usero04}. c-C$_3$H$_2$ and NH$_3$ were not detected in this survey, but were compiled from the literature: $^f$\, \citet{Nakajima11} and $^g$\,\citet{Ao2011}.
\end{list}{}{}

\end{table}

\begin{table}
\caption{Upper limits to the column densities of some undetected species.}
\centering
\begin{tabular}[!h]{lcccc} 
\hline
\hline
Molecule	& 	$N_{\rm mol}$	& 	Molecule	& 	$N_{\rm mol}$	\\
		&	cm$^{-2}$	&			&	cm$^{-2}$\\ 
\hline
H$_2$CO		&	$\le1.1\times10^{16}$ & H$_2$CS		&	$\le2.6\times10^{14}$\\
OCS		&	$\le5.4\times10^{14}$ & C$_2$S		&	$\le8.9\times10^{13}$ \\
CH$_2$NH	&	$\le5.4\times10^{14}$  &HOCO$^+$	& 	$\le7.0\times10^{13}$ \\ 
CH$_3$CCH	&	$\le4.5\times10^{14}$  &$^{13}$CS	&	$\le3.6\times10^{13}$  \\ 
SO$_2$		&	$\le3.6\times10^{14}$ & NH$_2$CN	&	$\le2.8\times10^{13}$   \\ 
c-C$_3$H	&	$\le2.9\times10^{14}$  \\ 
\hline
\end{tabular}

\begin{list}{}{}
\item[]See Sect.~\ref{sect.results} for details about the calculations.
\end{list}{}{}

\label{TableUpper}
\end{table}

\subsection{Isotopic line ratios}
\label{compratios}
\begin{table}
\caption{Molecular line ratios in NGC\,1068 and comparison with the isotopic ratios in other galaxies.}
\centering
\begin{tabular}[!h]{lcccc} 
\hline
\hline

 NGC\,1068	& GC	& M\,82		\\
$J-J'=1-0$	&			\\ 
\hline
 $^{12}$CO\,/\,$^{13}$CO\,=\,12.6	 \hspace{7 mm} \rdelim\}{5}{2mm} & 		\\ 
 H$^{12}$CO$^+$\,/\,H$^{13}$CO$^+$\,=\,17.1 	 &	   \\ 
HN$^{12}$C\,/\,HN$^{13}$C\,=\,23.3		& $^{12}$C\,/\,$^{13}$C$\sim20-25\,^g$ & $^{12}$C\,/\,$^{13}$C$>138\,^e$ \\  
 H$^{12}$CN\,/\,H$^{13}$CN\,=\,16.1	 &	\\
$^{12}$CN\,/\,$^{13}$CN	\,=\,49.1	&  \\ 
C$^{16}$O\,/\,C$^{18}$O\,=\,$ 176.8\,^a$		 & $^{16}$O\,/\,$^{18}$O\,=\,250$\,^c$ & $^{16}$O\,/\,$^{18}$O\,$>350\,^e$\\
 C$^{32}$S\,/\,C$^{34}$S\,=\,5.2$\,^b$		 & $^{32}$S\,/\,$^{34}$S$\sim22\,^d$ & $^{32}$S\,/\,$^{34}$S\,=\,$20\,^f$\\
\hline
\end{tabular}

\begin{list}{}{}
\item[]$a$: See details in Sect.~\ref{compratios} about how this value was calculated; $b$: Using data from the literature (see Table~\ref{TableNT} for references); $c$: \citet{Wilson94}; $d$: \citet{Frerking80}; $e$: \citet{Martin10}, using C$_2$H and its two C-isotopologues; $f$: \citet{Martin09a} and \citet{Aladro10}; $g$: \citet{Wannier80}. 

\end{list}{}{}

\label{ratios}
\end{table}

Isotopic ratios are commonly studied through the line ratios between different molecular species and their isotopic substitutions. However, molecular line ratios are often affected by opacity effects, selective photodissociation or/and isotopic fractionation, thus giving only lower limits to the actual values of isotopic ratios. Fractionation processes could affect the carbon isotopic molecular ratios (but not the oxygen and sulphur ones, \citealt{Wannier80,Woods09}), and might be more effective in regions with low kinetic temperatures (such as cold dark clouds, \citealt{Sakai10}), and may be not negligible for NGC\,1068 given the low temperatures traced by some molecules (see Sect.~\ref{temperatures}). On the other hand, in the presence of strong UV radiation, isotope-selective photoionization can also take place in the external regions of molecular clouds, where shelf-shielding and mutual shielding between different carbon isotopologues can occur \citep{Glassgold85}. This seems to have a negligible impact 
on the $^{12}$CO\,/\,$^{13}$CO ratio, although it may modify the C$^{16}$O\,/\,C$^{18}$O result \citep{Chu83,Wilson94}. UV fields may be strong in the nucleus of NGC\,1068 (see our modelling results in Sect.~\ref{modelling}) so, apart from our $^{12}$CO\,/\,$^{13}$CO, we cannot rule out that selective photoionization is affecting our lower limits to the carbon and oxygen isotopic ratios.
Finally, opacity effects clearly influence our derived $^{12}$C\,/\,$^{13}$C ratios, as they certainly affect the main isotopologues of CO, HCO$^+$ and HCN in NGC\,1068. On the contrary, $^{13}$C-bearing isotopic species are expected to be optically thin, or only moderately optically thick.

We used the detected transitions of CO, HCN, HNC, HCO$^+$, CN, and their $^{13}$C isotopologues to calculate a lower limit to the $^{12}$C\,/\,$^{13}$C ratio in the nucleus of NGC\,1068.  On the other hand, using the observed lines of C$^{16}$O, C$^{18}$O, C$^{32}$S, and C$^{34}$S, we derived limits to the $^{16}$O/$^{18}$O and $^{32}$S/$^{34}$S isotopic ratios. Our results are summarised in Table~\ref{ratios}. They were calculated by dividing the column densities listed in Table~\ref{TableNT}.

Our molecular line ratio $^{12}$CO\,$(1-0)$\,/\,$^{13}$CO$(1-0)=12.6$ is in accordance with previous results, where moderately optically thickness of $\tau \sim1-2$ for CO\,$(1-0)$ was found \citep{Young86,Papadopoulos96}. The HCO$^+(1-0)$\,/\,H$^{13}$CO$^+(1-0)$, HCN$(1-0)$\,/\,H$^{13}$CN$(1-0)$, and HNC$(1-0)$\,/\,HN$^{13}$C$(1-0)$ ratios give similar values, of  17.1, 16.1 and 23.3, respectively. Among all, CN is the species giving the highest lower limit to the $^{12}$C\,/\,$^{13}$C ratio, of 49.1.
 
Using the C$^{16}$O\,$(1-0)$ and C$^{18}$O\,$(1-0)$ column densities, we obtained $^{16}$O\,/\,$^{18}$O\,=\,50.0. This low value points to the opacity of both C$^{16}$O\,$(1-0)$ and C$^{18}$O\,$(1-0)$ lines (having the later a $\tau \ge 1$, \citealt{Papadopoulos99}). However, if we calculate this ratio using  $^{16}$O\,/\,$^{18}$O\,=\,$\frac{I\,({^{13}C^{16}O})}{I\,({^{12}C^{18}O})} \times \frac{^{12}C}{^{13}C}$, and the $^{12}$C\,/$^{13}$C\,=\,49.1 obtained from CN, we obtain a higher lower limit to  the $^{16}$O\,/\,$^{18}$O ratio of 176.8.

Unfortunately, our CS\,$(2-1)$ line is contaminated by the emission of $^{13}$CO\,$(1-0)$ coming from the image band (see Sect.~\ref{analysis}). Therefore, we used the CS\,$(3-2)$ and C$^{34}$S\,$(3-2)$ detections by \citet{Martin09a} and \citet{Bayet09b} to calculate  a limit to $^{32}$S\,/\,$^{34}$S, which gives a result of 5.2.

\section{Modelling of the data}
\label{modelling}

 \begin{table*}
\caption{Main parameters of the UCL\_CHEM models}
\centering
\begin{tabular}[!h]{lcccccc} 
\hline
Parameter	& Model $a$	 &	Model $b$	& Model $c$ & Model $d$	& Model $e$\\
\hline
Temperature (phase II)$^\dagger$	& 200/100\,K & 300/100\,K & 250/300\,K & 350/300\,K & *\\
Visual extinction	& 2 \& 10\,mag &  2 \& 10\,mag & 2 \& 10\,mag & 2 \& 10\,mag & *\\
External UV radiation intensity	& 1 Habing & 1000 Habing & 1 Habing & 1000 Habing &  1 Habing\\
Cosmic ray ionization rate ($\zeta$)&	1.3$\times$10$^{-17}$\,s$^{-1}$ & 1.3$\times$10$^{-17}$\,s$^{-1}$ & 1.3$\times$10$^{-15}$\,s$^{-1}$ & 1.3$\times$10$^{-15}$\,s$^{-1}$ &  1.3$\times$10$^{-15}$\,s$^{-1}$s\\
 \hline
 \end{tabular}

 \label{UCLCHEM}
\begin{list}{}{}
\item[]Models $a$ and $b$, characterized by a low  $\zeta$,  aim to explore how the chemistry changes as a function of the FUV intensity.
\item[]Models $c$ and $d$, characterized by a high $\zeta$, aim to explore how the chemistry of a XDR-like environment changes as a function of FUV intensity.
\item[]$^*$ In Model $e$ the UCL\_CHEM was coupled with the parametric shock model developed by \citet{Izaskun09}, and in this case  the temperature in phase II and the visual extinction vary as the shock passes through the gas.
\item[]$^\dagger$ First value corresponds to A$_{\rm V}=2$\,mag and second to A$_{\rm V}=10$\,mag.

\end{list}{}{}

 \end{table*}

To model the molecular line emissions presented in Sect. \ref{sect.results}, we used the time and depth dependent chemical model UCL\_CHEM \citep{Viti99,Viti04}. The molecular emission from the AGN of NGC\,1068 arises from an ensemble of gas components ranging in density and temperature by several orders of magnitude, representing various stages of the ISM, from diffuse to dense gas. Hence, quantitatively modelling the observed line emission would require some knowledge of the geometry as well as the dynamics of the different gas components. Instead chemical and photon-dominated region (PDR) models can be used to qualitatively determine the origin of each molecule and give an indication of the relative contribution among the different gas components. 

The UCL\_CHEM was used to first simulate the formation of a dark core by collapsing an atomic diffuse gas from a density of 100\,cm$^{-3}$ to a final density of $10^5$\,cm$^{-3}$. During the collapse atoms and molecules freeze onto the dust grains forming icy mantles. Both gas and surface chemistries are self-consistently computed. Once the gas is in dynamical equilibrium again, UCL\_CHEM compiles the chemical evolution of the gas and the dust after a burst of star formation has occurred. So, while the temperature during the first phase of the modelling was kept to 10K, during the second phase it is increased to 100-350 K  and the icy mantles are evaporated. The chemical evolution of the gas is then followed for 10$^7$ years. In both phases of the UCL\_CHEM model the chemical network is based on more than 2345 chemical reactions involving 205
species of which 51 are surface species. The gas phase reactions were adopted from the UMIST data base \citep{Millar97, LeTeuff00, Woodall07}. All the molecules so far detected in galactic hot cores
(e.g. \citealt{Blake87,Millar98,Hatchell98,Fontani07}) have been included in this chemistry network together with all the important and familiar molecular species that are known from many other studies to play a significant role in the chemistry of interstellar gas. The surface reactions included in this model are assumed to be mainly hydrogenation reactions, allowing chemical saturation where this is possible. 

The adopted values for the free parameters of the UCL\_CHEM model are listed in Table~\ref{UCLCHEM}. Parameters not mentioned in this table have been kept to their standard (Milky Way) values. We used a final density of n$\rm _H$\,=\,$10^5$\,cm$^{-2}$ and a metallicity value of z\,=\,1.056\,z$_\odot$, based on previous studies of NGC\,1068 (e.g. \citealt{Zaritsky94,Tacconi94,Kamenetzky11}). The initial elemental abundances (C/H, O/H, N/H, S/H, He/H, Mg/H and Si/H) used in our study are those corresponding to extragalactic environments (as described in \citealt{Bayet08}), scaled to the NGC\,1068 metallicity.

To sample the likely co-existing conditions that molecules are experiencing in the
nucleus of NGC 1068, we ran five models. The first four ($a$, $b$, $c$, and $d$; see Table~\ref{UCLCHEM} for details),  allowed us to investigate the response of the chemistry to the changes in FUV radiation field and  cosmic ray ionization rate, $\zeta$. High cosmic ray ionization rates can be used to qualitatively simulate XDR-like environments (see e.g. \citealt{Bayet09a}). In fact, only few species (OH$^+$, OH, H$_2$O, H$_3$O$^+$) show clear differences in abundances between regions pervaded by cosmic rays and X-rays  \citep{Meijerink11}. Since none of these molecules were observed in our survey, for the purpose of the modelling we assume that both enhanced $\zeta$ and X-rays leave similar fingerprints in the ISM. 

Our models $c$ and $d$, which are characterized by a high $\zeta$ of 1000 times the canonical value of the Milky Way, aim to reproduce chemical conditions similar to the NGC\,1068 nucleus. On the other hand, model $e$ aims to study the possible influence of C-shocks, as it was recently claimed that they may be affecting the gas in the CND of NGC\,1068 \citep{Burillo10,Krips11, Kamenetzky11}. To do this, we used the initial conditions of model $c$ and ran a version of UCL\_CHEM coupled with the parametric shock model developed by \citet{Izaskun09}, which considers a plane-parallel C-Shock propagating with a velocity $v$ (we set $v=40$\,kms$^{-1}$) through the ambient medium. Details of this version of UCL\_CHEM can be found in \citet{Viti11}. The influence of shocks in the NGC\,1068 nucleus is further discussed in Sect.~\ref{comp}.

Using the results of the models, we calculated the column densities of the seventeen species detected in our survey (carbon, oxygen and sulphur isotopologues are not included) at a representative time of $10^5$\,yr (but note that chemistry has not necessarily reached steady state by then), using: \\
\begin{equation}
\label{eq1}
N_{\rm mol} = X_{\rm mol} \times A_{\rm V} \times 1.6\times10^{21}
\end{equation}
where $X_{\rm mol}$ is the fractional abundance of the molecule, $A_{\rm V}$ is the visual extinction,
and $1.6\times10^{21}$\,cm$^{-2}$ is the column density of hydrogen at a visual extinction of one
magnitude. Note that the formulation above simply give an `on the spot' approximation of the column density. We set $N_{\rm mol}=10^{12}$\,cm$^{-2}$ as the minimum theoretical column density to consider a species detectable. Below this value we do not take into account the results of the models. We included, however, some species that were not detected in our survey (CH$_3$CCH, H$_2$CO, SO$_2$, H$_2$CS), but whose response to UV fields, cosmic rays and shocks may help us understand the differences in chemistry between AGNs and starburst galaxies.

\subsection{Origin of the molecules}

Disentangling the origin of extragalactic molecular emission is not trivial, as often the same molecules may arise from very different gas conditions (e.g. Bayet et al. 2008, 2009). For the purpose of this analysis, we  considered a two-components model: molecular emission arising from the outer layers of molecular clouds, which are more exposed to external radiation fields that form PDRs; and emission arising from denser gas, closer to the centre of the clouds. These two gas components are represented in our model by  visual extinctions of $A_{\rm V}=2$\,mag (PDRs) and 10\,mag (dense gas), similarly to previous studies \citep{Bayet08, Bayet09a,Meijerink05, Spaans05}. We note that above a certain critical visual extinction ($\sim$ 2.5-3 mag, depending on the radiation field) the chemistry is no longer affected by UV radiation and therefore any $A_{\rm V}$ above that could have been chosen to represent the denser, shielded gas component. Hence, increasing or decreasing the $A_{\rm V}$ would simply imply a 
variation in the estimated column densities by the same factor (see Eq.~\ref{eq1}).

In Table~\ref{origin}, we list all molecules and we classify them as `tracers' of a particular gas component for each model. We found that C$_2$H and CN always trace PDRs and dense gas independently of the strength of both UV fields and CRs (i.e. according to all the models). On the other hand, CO, HCN, CS, SiO, CH$_3$CN, HNC, SO, CH$_3$CCH, H$_2$CO, and SO$_2$ are invariably found in dense gas regions. We note that many molecules can trace PDRs as well as the denser, shielded regions, depending on the strength of UV fields and CRs adopted (Table~\ref{origin}), highlighting once more (e.g. Bayet et al. 2009) that one can not easily isolate PDR-tracers as templates for modelling galaxies. These results generally agree with previous observational and theoretical studies (e.g. \citealt{Huggins84b,Sternberg95,Bayet08,Bayet09a,Aladro10}).

\subsection{Influence of UV fields and cosmic rays on molecules}

Most molecular species are sensitive to the presence of external fields. We checked the variations of their column densities as a function of UV and CRs strengths (i.e. for the models $a$, $b$, $c$, and $d$). A summary of the findings is shown in Table~\ref{resumen}. While the column density of HOC$^+$ is enhanced by the presence of UV radiation, many other species are easily dissociated by UV fields in the external layers of the molecular clouds (such as CO, HCN, or NS).  On the contrary, we found that methanol (CH$_3$OH) is clearly destroyed by cosmic rays, while the production of several other species is favoured by the presence of CR photons in the ISM (e.g. SiO, CN, or N$_2$H$^+$). Of particular interest is SiO, commonly considered as a shock tracer, but whose enhanced abundance in NGC\,1068 can also be related to high rates of X-rays \citep{Usero04,Burillo10} and cosmic rays (this work). Finally, only C$_2$H shows similar column densities (within one order of magnitude of variation) for all the models.

Our results agree well with those of \citet{Bayet08}, who also used the UCL\_CHEM model to find suitable molecular tracers of hot cores under different physical conditions. In general, the trends of chemical abundances with respect to $\zeta$ variations found  by \citet{Bayet08} are similar to what is shown in our Table~\ref{resumen}, with few exceptions, being CS the most contrasted species between both works. For this species  \citet{Bayet08} predicted a decrease of its abundance with increased $\zeta$, while our models predict a slight increase. This may be explained by the different physical parameters used in both works (other $\zeta$ strengths, initial and final densities, visual extinction and final temperatures).

On the other hand, the comparison of our models with those that explored the PDR chemistry in a variety of extragalactic environments \citep{Bayet09a}, leads to big discrepancies regarding the response of the molecular abundances as a function of $\zeta$ and FUV radiation fields. For example, \citet{Bayet09a} deduced that CN and HCO$^+$ are not affected by the variations of $\zeta$, while our results predict enhancements of those species. Also, \citet{Bayet09a} found that CO, HNC, HCN and H$_2$CO are not influenced by changes of the FUV radiation field, whereas our results indicate that the abundances of these molecules are reduced by at least one order of magnitude when the FUV field increases. These disagreements are mainly due to the fact that \citet{Bayet09a} used the UCL\_PDR model where, unlike the UCL\_CHEM code used here, the depletion of atoms and molecules on to grains, and the subsequent surface chemistry, is not included. In other words, our present modelling assumes that even the gas 
affected by UV photons has undergone sufficient processing involving gas-grain interactions.

\begin{table}
\caption{Origin of the molecules according to each model}
\centering
\begin{tabular}[!h]{lcccccccc} 
\hline
\hline
Molecule & Model $a$ &  Model $b$ & Model $c$ & Model $d$ \\
\hline
CO & P\& D & D  & P\& D   & D\\
C$_2$H &	P\& D &P\& D &P\& D &P\& D \\
CN 	&	P\& D &P\& D &P\& D &P\& D \\
HCN &	 P\& D & D  & P\& D   & D\\
CS& P\& D & D  & P\& D   & D	\\
HCO$^+$& - & -& P\& D   & D\\
HNC &P\& D & D  & P\& D   & D\\
HCO &P\& D & - & P\& D   & -	\\
N$_2$H$^+$ & - &- & P\& D   & D	\\
SiO &D   & D   &P\& D   & D	 \\
CH$_3$CN &D   & D   &P\& D   & D\\
NS & - &- & P\& D   & D		\\
H$_2$CO &D   & D   &P\& D   & D	\\
HC$_3$N & D &- &-&-\\
 CH$_3$OH &D &- &-&-	\\
SO &D & D &D&D\\
CH$_3$CCH &D & D &D&D\\
SO$_2$ 	&D & D &D&D	\\
H$_2$CS	&D & D & -&-	\\
HOC$^+$ &-&-&-&- \\
\hline
\end{tabular}

\begin{list}{}{}
\item[] $P$ stands for PDR emission, while $D$ stands for dense gas emission (evaluated at visual extinctions of 2\,mag and 10\,mag respectively). $P\&D$ indicates that the molecule may arise from both molecular cloud regions.`-' means that the model give a column density below $10^{12}$\,cm$^{-2}$ (theoretical detectable limit), and thus the result is not taken into account.

\end{list}{}{}

\label{origin}
\end{table}





\begin{table}
\caption{Molecular trends with UV fields and cosmic rays}
\centering
\begin{tabular}[!h]{llcccccccc} 
\hline
\hline

Easily dissociated by UV fields &  CO, CS, HCN, CH$_3$OH,\\
				& HCO, SO, NS, HNC, SO$_2$,		\\
				&  N$_2$H$^+$, SiO, CH$_3$CN, H$_2$CS,	\\
				& H$_2$CO, HC$_3$N, CH$_3$CCH	\\
Enhanced by UV fields		& HOC$^+$, 		\\
Easily dissociated by cosmic rays & CH$_3$OH, HC$_3$N		\\
Enhanced by cosmic rays		& CN, HCN, HCO$^+$, HNC, SO,	\\
				& NS, N$_2$H$^+$, SiO, SO$_2$, HOC$^+$	\\
Insensitive			& C$_2$H \\
\hline
\end{tabular}


\label{resumen}
\end{table}

\subsection{Comparison with the observations}
\label{comp}

By considering a multi-component model we succeed in qualitatively reproducing the observed column densities of almost all the detected molecules in this survey within one order of magnitude, being HOC$^+$ the only exception. We consider this a good match, as the column densities have relevant uncertainties due to the assumption of LTE conditions. In general, the observations are better reproduced by the models which consider a high cosmic ray ionization rate, but low FUV radiation field (model $c$), or high values of both parameters (model $d$). Once more, this result supports the use of high $\zeta$ rates in time- and depth-dependent chemical models as a good approximation of XDR-like environments.

However, as expected, the scenario depicted by the models reflects clear differences in the origin and timescales of the species, as well as highlights some degeneracies; nevertheless sometimes the same degeneracies can emphasise the prevalence of a particular energetic process. For example CO, ubiquitously tracing gas at a wide range of densities, is well matched by all the models, yet models with high and low radiation field, as long as the cosmic ray ionization rate is high, match the observations better. On the other hand, methanol is much better matched by a model where both radiation field and cosmic ray ionization rates are low: we interpret this `mismatch' in physical conditions between CO and CH$_3$OH as an indication that the two species, on average, arise from different gas; methanol clearly seem to arise from regions where the cosmic ray ionization rate is close to standard, probably not representative of the average cosmic ray ionization rate of the galaxy. This rate is in fact well 
known to be variable within our own galaxy \citep{Dalgarno06}, and it may well vary in NGC1068 too. On the other hand, the HCO$^+$ observed abundance is only matched if the cosmic ray ionization rate is high (i.e. Models $c$ and $d$), independently of the FUV field strength.

Unlike for cosmic rays, there are very few species that help us to determine the average UV field(s) strength in NGC1068. Of particular interest is HCO, which seems to get closer to the observed value only in models where the UV field is low, while there is no molecule that {\it needs} a high radiation field in order to match the observations. While this does not exclude a high radiation field for this galaxy, especially near the nucleus, it seems unlikely that a rich chemistry comprising of species such as HCO would survive if the average radiation field were to be much higher than the canonical interstellar one.

On the other hand, C$_2$H is better reproduced in a shocked environment, as C-type shocks enhance its abundance up to the observational value. On the contrary, H$_2$CO seems to be destroyed by C-shocks. This, as well as strong UV fields, might be the reason for its non-detection in our survey. Apart from C$_2$H and H$_2$CO, the presence of shocks does not greatly affect the abundances of our observed species. In particular we note that the SiO column density does not change if shocks are present, suggesting that its enhanced abundance is due to X-rays/cosmic rays gas processing instead of shocks. Hence, although they might be present, our models do not require the presence of strong shocks in the CND of NGC\,1068.  It is worth underlining here, however, that our models only include a treatment for C-type shocks: in these models, once the dense gas reaches its final pre-shocked density (i.e. at the end of Phase I), the gas and dust temperatures are already high enough (250K) that many exothermic reactions are relatively efficient. The maximum temperature reached due to the shock is only few hundred kelvin higher than the initial temperature, and moreover it does not last for long ($\sim$ 100-200 years) before the gas starts cooling down. Hence, chemically  there is not much difference between this scenario and the typical `hot core' scenario. The introduction of J-type shocks would significantly alter the chemistry.

Finally, it is interesting to note that our models show different recycling times among molecules. This behaviour seems to be related to the nature of the species rather than a dependence on the radiation field and cosmic rays strength. For example, the column densities of some species, such as CH$_3$OH and NS, are better matched at early times ($\sim10^3$\,yr), since their abundances quickly drop at later times by at least three orders of magnitude. Similarly, the column densities of some important undetected species (e.g. CH$_3$CCH) drop dramatically below the detectable levels at later times (i.e. $\sim10^5$\,yr). On the other hand, CO, HCO$^+$ and HOC$^+$ always maintain constant values.

\section{Comparison with other galactic circumnuclear regions}
\label{comparison}

\subsection{Isotopic ratios}

For comparison purposes, Table~\ref{ratios} lists the isotopic line ratios obtained in the Galactic centre (GC) and the inner few hundred parsecs of the starburst galaxy M\,82. As discussed in Sect.~\ref{compratios}, our upper limits to the $^{12}$C\,/\,$^{13}$C ratio in NGC\,1068 are certainly affected by opacity effects. Also, selective photoionization and chemical fractionation could be altering our carbon and oxygen isotopic ratios in NGC\,1068. These effects might partially explain the isotopic differences among galaxies. 

The most stringent lower limit to the $^{12}$C\,/\,$^{13}$C ratio obtained in M\,82 so far ($>138$, \citealt{Martin10}) is more than two times higher than our best value in NGC\,1068. However, we note that the   M\,82 ratio was calculated from C$_2$H, while for NGC\,1068 we are using our CN results. Both molecules can be tracing different gas phases in the galaxies, so it is not surprising that they show distinct values. If we compare the carbon isotopic ratios of both galaxies using CO and HCO$^+$ \citep{HenkelMauers93}, the results are similar.

On the other hand, the difference in carbon isotopic ratios between galaxy nuclei may also be related to the degree of processed material. $^{12}$C and $^{13}$C are products of primary and secondary nuclear processing in stars respectively, and thus their ratio depends on the nucleosynthesis of stars, on stellar evolution, and therefore on the star formation history of the galaxy. The stellar population age in the centre of NGC\,1068 (200-300\,Myr, \citealt{Davies07}) is much older than that of M\,82 ($\sim15$\,Myr, \citealt{Konstantopoulos09}). Primary-to-secondary isotopic ratios are predicted to decrease with time, so a lower $^{12}$C\,/\,$^{13}$C ratio (as well as $^{16}$O\,/\,$^{18}$O) is expected in NGC\,1068 than in M\,82. On the other hand, both the Galactic centre and the NGC\,1068 nucleus have low carbon isotopic ratios as compared to M\,82 (Table~\ref{ratios}). This implies that both nuclei have a high chemical processing degree.

Our  limit to the  $^{32}$S\,/\,$^{34}$S in NGC\,1068 is a factor of four lower than the ratios in the GC and M\,82. In the Milky Way, it was observed that $^{32}$S\,/\,$^{34}$S does not strongly vary with time, although increases with the galacticentric distance, following the  $^{12}$C\,/\,$^{13}$C gradient of the galactic disk \citep{Frerking80,Wilson94,Chin96}. Also, it was proposed that this ratio depends on the stellar population, so that environments with more massive stars and supernova explosions (which is the case of NGC\,1068 as compared to the GC and M\,82) have lower $^{32}$S\,/\,$^{34}$S.

\begin{figure*}[h!]
\begin{center}
\includegraphics[angle=0,width=\textwidth]{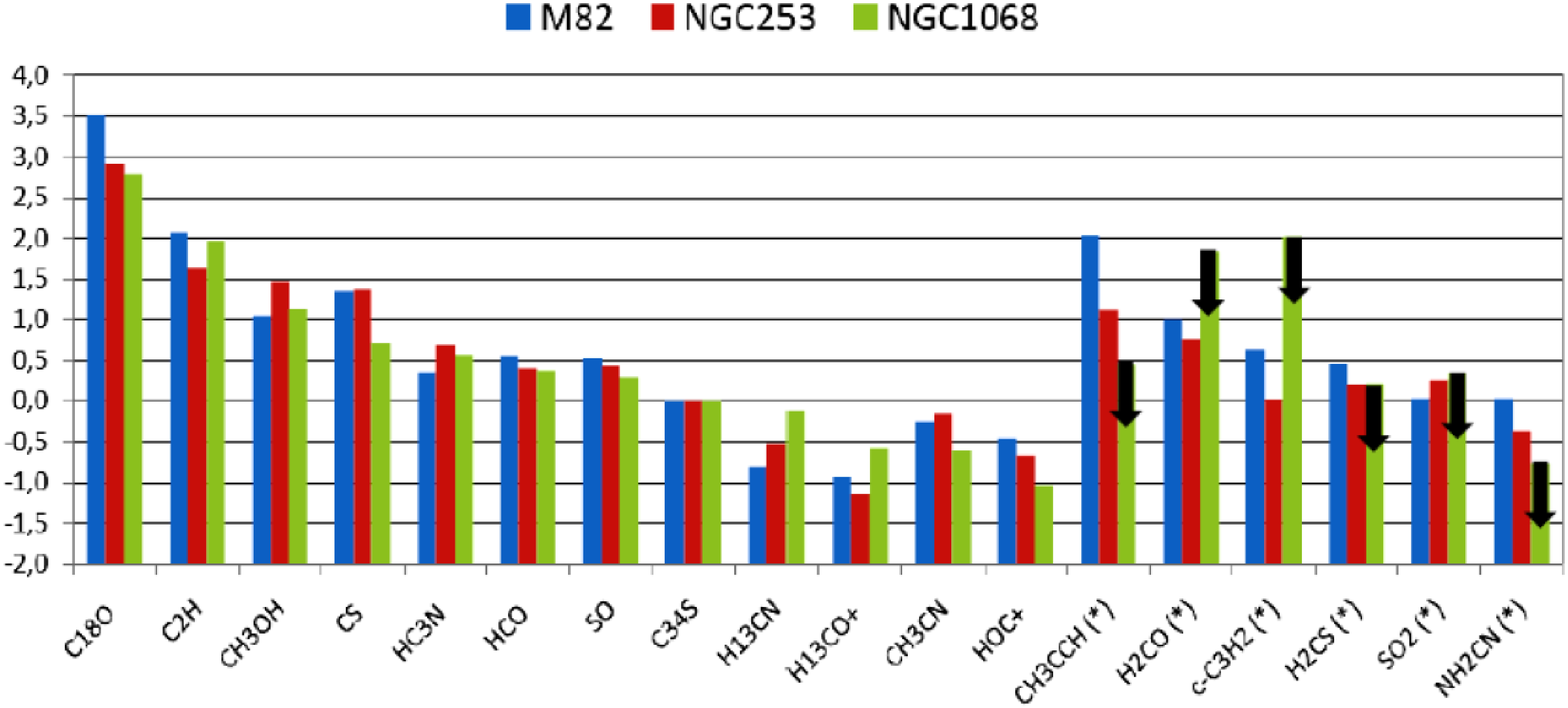}
\caption{Comparison of the fractional abundances among the two starbust galaxies M\,82 (blue) and NGC\,253 (red), and the AGN NGC\,1068 (green). Y-axis is $\rm log_{10}(N_{\rm mol}/N_{\rm {C^{34}S}}$). Molecules marked with an * were not detected in our NGC\,1068 survey; arrows indicate their upper limits to the fractional abundances.}
\label{Abundances}
\end{center}
\end{figure*}

\subsection{Molecular abundances}
M\,82 and NGC\,253 are the starburst galaxies with the best known chemical compositions so far. Like NGC\,1068, both have been the targets of unbiased molecular line surveys toward their nuclei \citep{Martin06b, Aladro11}. The starburst in the centre of M\,82 is old, with an average stellar population age of $\sim$10\,Myr \citep{Konstantopoulos09}. This, together with a high supernova rate, creates strong UV fields, so that PDRs dominate its chemistry. On the other hand, NGC\,253 is claimed to be in an earlier stage of starburst evolution \citep{Martin09a,Aladro10}, with younger stellar populations in its nucleus ($\sim6$\,Myr, \citealt{Onti09}) where the PDRs, although present, do not characterise its ISM. Instead, the nucleus of NGC\,253 show that large-scale shocks are the dominant process driving the molecular abundances.

Figure~\ref{Abundances} shows the abundances of all species detected by single-dish mm/sub-mm telescopes in the three galaxies, as well as other species that were identified in both starburst galaxies but remain undetected in NGC\,1068 (for them we used the upper limits to the column densities listed in Table{~\ref{TableUpper}). This figure aims to highlight the chemical differences between AGN-dominated and starburst-dominated nuclei. The fractional abundances were calculated with respect to C$^{34}$S, which is significantly less affected by opacity effects than CS, and whose abundance does not seem to be significantly affected by the type of nuclear activity \citep{Martin09a}. 

We note that while M\,82 and NGC\,253 are located at about the same distance (3.6 and 3.9\,Mpc respectively, \citealt{Freedman94,Karachentsev03}), NGC\,1068 lies four times further. However, while the comparison of the column densities in Fig.~\ref{Abundances} would be biased by the different distances to the galaxies, this should not affect the comparison of the fractional abundances.

Among all the species compared, only the ratios H$^{13}$CN/C$^{34}$S and H$^{13}$CO$^+$/C$^{34}$S are higher in NGC\,1068 than in the starburst galaxies, being between two and five times more abundant in the AGN. Conversely, the fractional abundances of C$_2$H, CH$_3$OH, HC$_3$N, HCO and SO show similar values in both types of nuclei, with differences within a factor of two among the galaxies. This suggests that these species are not especially affected by the different nuclear activities in starburst galaxies and AGN, as previously suggested by \citet{Nakajima11}. A third group is formed by those species whose fractional abundances are clearly higher in both M\,82 and NGC\,253 than in NGC\,1068. Examples are CH$_3$CN/C$^{34}$S, or HOC$^+$/C$^{34}$S, which show differences in the abundances between three and five times. It is remarkable that these molecules have very similar abundances in M\,82 and NGC\,253. However, the biggest contrasts are given by those molecules that have been clearly detected in both 
starburst galaxies, but not in our NGC\,1068 survey. Examples are CH$_3$CCH, H$_2$CO, c-C$_3$H$_2$, H$_2$CS, SO$_2$ and NH$_2$CN. We note that propyne (CH$_3$CCH) and formaldehyde (H$_2$CO) are abundant species in starburst galaxies \citep{Muhle07,Aladro10}. Unlike H$_2$CO, we obtained a stringent upper limit to the  CH$_3$CCH column density in NGC\,1068 (Table~\ref{TableUpper}), which makes it the most contrasted species of our survey when comparing starburst galaxies and AGN (Fig.~\ref{Abundances}). We found that CH$_3$CCH/C$^{34}$S is $\sim$40 and $\sim$5 times more abundant in M\,82 and NGC\,253, respectively, than in NGC\,1068. This strong variation in the CH$_3$CCH fractional abundances is likely to be intrinsically related to the different nuclear activities in galaxies, rather than being due to other factors, such as different opacities (due to its K-ladder structure the transitions of this molecule are expected to be optically thin), or kinetic temperature of the gas (as we do not find a strong (
anti-)
correlation of its abundances with the $T_{\rm kin}$ of the three galaxies).



\section{Conclusions}
\label{conclusions}

We used the IRAM-30\,m telescope to carry out a spectral line survey (from 86.2\,GHz to 115.6\,GHz) towards the nucleus of the AGN galaxy NGC\,1068. We detected a total of 35 spectral lines corresponding to 17 molecular species. Seven molecules, or their $^{13}$C isotopologues, are detected for the first time in this galaxy, namely HC$_3$N, SO, N$_2$H$^+$, CH$_3$CN, NS, $^{13}$CN, and HN$^{13}$C. We note that some molecules that are abundant in other galaxies, such as H$_2$CO or CH$_3$CCH, are not detected in NGC\,1068. We calculated the column densities of the identified species assuming local thermodynamic equilibrium and optically thin emission. However, some intense lines such as $J-J'=1-0 $ CO, HCO$^+$, HCN or HNC, are affected by optical thickness (and might be also altered by chemical fractionation and isotope-selective photoionization effects), so their column densities are lower limits to the actual values.

Using several lines from our survey we obtained the isotopic abundance ratios $^{12}$C\,/\,$^{13}$C$\,=49$,  $^{16}$O\,/\,$^{18}$O\,$\ge 177$ and $^{32}$S\,/\,$^{34}$S\,$=5$ for NGC\,1068. These isotopic ratios significantly differ from those of the Galactic centre and starburst galaxies. Apart from the fact that the opacities of the transitions change from source to source, this reflects the different star formation histories of the galaxies.

We ran chemical models to determine the origin of the species detected in our survey, as well as to test the influence of UV fields, cosmic rays, and shocks on their abundances. We used a high rate of cosmic ray ionization rate to qualitatively simulate a XDR-like environment. We found that the assumption of some molecules being PDR tracers or dense gas tracers should be done carefully, and only if the physical conditions of the gas are known, since under certain circumstances a PDR tracer may be also arising from dense gas regions and viceversa. Our models indicate that a high cosmic ray ionization rate is present in the nucleus of NGC\,1068, albeit a combination of a high cosmic ray ionization rate and UV fiels may also explain the observed abundances in the galaxy. On the contrary, shocks do not seem to be necessary to explain most of the molecular abundances. The chemical models applied to NGC\,1068 predict that while UV radiation easily dissociate a large number of molecules (e.g. HCN, CH$_3$OH, HNC, or 
CH$_3$CCH), cosmic rays enhance the abundances of a good number of others, such as CN, SiO, N$_2$H$^+$, NS, or HCN.

We aimed to find the molecules that are more sensitive to the different power sources of galaxies nuclei (starbursts and active galactic nuclei). We carried out a detailed comparison of the chemical composition of the ISM in the NGC\,1068 nucleus and those of the two starbursts galaxies M\,82 and NGC\,253. Examples of differing ratios are H$^{13}$CN/C$^{34}$S, which is $\sim$3-5 times more abundant in  NGC\,1068 than in starburst galaxies, and CH$_3$CCH/C$^{34}$S, which is, at least,  $\sim$5-40 times more abundant in starburst galaxies than in NGC\,1068.

\begin{acknowledgements}
We thank the IRAM staff for their help with the observations. This work was supported by the Science \& Technology Facilities Council [ST/F501761/1] and also the Spanish Ministerio de Ciencia e Innovaci\'on under the projects ESP2007-65812-C02-01 and AYA 2010-21697-C05-01. SM acknowledges the co-funding of this work under the Marie Curie Actions of the European Commission (FP7-COFUND).
\end{acknowledgements}



\clearpage

\begin{appendix}
\section{Comments on individual molecules}
\label{sect.molecules}
We have detected seventeen molecular species, and seven carbon, sulphur, and oxygen isotopic substitutions in the 3\,mm atmospheric window, between 86.2\,GHz and 115.6\,GHz. Here, we detail each observed transitions, the Gaussian profile fitting, cases of blended lines and other special cases. In order to identify the molecules, we started from the premise that only the transitions having $E_{\rm low}<$100\,cm$^{-1}$ and log($\int$ $T_{\rm{MB}}\,\rm dv)>-6$\,nm$^2$\,MHz contribute to the observed lines. After checking this initial condition, we ensured that the spectral features that not fulfil these conditions can be considered negligible. 

NGC\,1068 has a heliocentric systemic velocity of $v_{sys}=1137$\,km\,s$^{-1}$ (value taken from NASA/IPAC Extragalactic Database - NED). Based on the values obtained from the Gaussian fits, we have chosen an average value of 1100\,km\,s$^{-1}$ for fixing this parameter when needed. We found two different line widths for the detected molecules in NGC\,1068. In general, strong lines have values in the range FWHM$\sim[230-260]$\,km\,s$^{-1}$, while faint lines have line widths of about $\sim[120-190]$\,km\,s$^{-1}$. The cases where it was necessary to fix this parameter are commented below, since the value depends on each species.

\begin{itemize}

\item{\bf Hydrogen cyanide -  HCN and H$^{13}$CN}\\
We detected the H$^{13}$CN\,$(1-0)$ transition at 86.3\,GHz. It is formed by three hyperfine spectral features, although the structure of the line is not resolved due to the broad line widths of NGC\,1068. To take into account the contribution of each hyperfine feature we have performed a synthetic Gaussian fit (details about this method can be seen in \citealt{Martin10,Aladro11}). The resulting values, shown in Table~\ref{TableGauss}, correspond to the main component of the group. This HCN carbon isotopologue is detected for the first time in this galaxy.

HCN\,$(1-0)$ is identified at 88.6\,GHz. In a similar way as for H$^{13}$CN, we fitted a synthetic Gaussian profile leaving the linewidth and line position free. 

\item{\bf Hydrogen isocyanide - HNC and HN$^{13}$C}\\
One strong line of hydrogen isocyanide was detected at 90.7\,GHz. The telescope beam possibly picked up some emission coming from the starburst ring that surrounds the CND of NGC\,1068. We complemented this transition with the HNC\,$(4-3)$ line observed by \citet{Perez09}.

The HN$^{13}$C$(1-0)$ transition at 87.1\,GHz has been tentatively detected for the first time in this galaxy. Because it is quite faint, the integrated intensity resulting from the Gaussian fit is highly affected by the baseline. This is reflected in the column density derived from the rotational diagram, which has a high associated error.

\item{\bf Ethynyl - C$_2$H}\\
C$_2$H\,$(1-0)$, formed by six hyperfine features, is detected at 87.3\,GHz. We  performed a synthetic Gaussian fit to calculate the contribution of each feature. Because we only have one detected transition of this species, we calculated its column density fixing the rotational temperature to $10\pm5$\,K. This is the fourth most abundant molecule after CO and its carbon and oxygen isotopologues.

\item{\bf Isocyanic acid - HNCO}\\
Four groups of transitions were detected at 87.9\,GHz, 109.5\,GHz, 109.9\,GHz and 110.3\,GHz. They are composed by 5 features each. The line at 109.9\,GHz is seen as a bump at the higher frequencies of C$^{18}$O, although both can be easily separated. The transition at 110.3\,GHz is very close to $^{13}$CO and blended with CH$_3$CN$(6_{4,0}-5_{4,0})$. We performed a synthetic Gaussian fit to all these lines, including other two HNCO transitions that are too faint to be detected, but that fulfil the conditions mentioned at the beginning of this appendix ($E_{low}<$100\,cm$^{-1}$ and log($\int$ $T_{\rm{MB}}\,\rm dv)>-6$\,nm$^2$\,MHz). In the Gaussian fitting we fixed the linewidth to 230\,km\,s$^{-1}$ and the line position to 1100\,km\,s$^{-1}$. The integrated area of our HNCO\,$(5_{0,5}-4_{0,4})$ line is consistent with the values obtained by \citet{Martin09a}. However, from the Boltzmann diagram we obtain a rotational temperature of $T_{\rm rot}\sim$ 30\,K, which is almost 5 times higher than the one 
obtained by \citet{Martin09a}. This difference is due to the transitions with higher energy levels we included. 

\item{\bf Oxomethylium - HCO$^+$ and H$^{13}$CO$^+$}\\
The strong line of HCO$^+\,(1-0)$ at 89.2\,GHz is the only detected transition in our line survey, and   shows some emission coming from the starburst ring. We also used the HCO$^+(3-2)$ line observed at 267.6\,GHz by \citet{Krips08}. The H$^{13}$CO$^+(1-0)$ line is detected at 86.8\,GHz, although it is blended with HCO$(1_{1,0}-0_{1,0})$ and SiO$(2-1)$. To disentangle the contribution of each molecule, we performed synthetic Gaussian fits to the three lines. We first fitted H$^{13}$CO$^+(1-0)$ fixing the position to 1100\,km\,s$^{-1}$ and the linewidth to 240\,km\,s$^{-1}$. Then, we fitted the other two molecules (see Fig.~\ref{SiO}).

\item{\bf Oxomethyl - HCO}\\
Only one molecular transition of this species has been detected along the survey. It is blended with H$^{13}$CO$^+\,(1-0)$ and SiO\,$(2-1)$ (see Figs.~\ref{Survey123} and ~\ref{SiO}). HCO\,$(1_{1,0}-0_{1,0})$ is located at 86.7\,GHz, and contains four features. After subtracting the oxomethylium Gaussian fit, we over-fitted a Gaussian profile to the HCO features with a fixed position of 1100\,km\,s$^{-1}$ and a linewidth of 240\,km\,s$^{-1}$.

\item{\bf Silicon monoxide - SiO}\\
As commented before, SiO\,$(2-1)$ at 86.8\,GHz is blended with H$^{13}$CO$^+(1-0)$ and HCO$(1_{1,0}-0_{1,0})$. After subtracting the Gaussian fits of the other two molecules, we fitted another synthetic Gaussian profile to the residuals, fixing only the position to 1100\,km\,s$^{-1}$ (see Fig.~\ref{SiO}). This is the strongest molecule of the three blended lines.

\begin{figure}
\begin{center}
\includegraphics[width=0.5\textwidth]{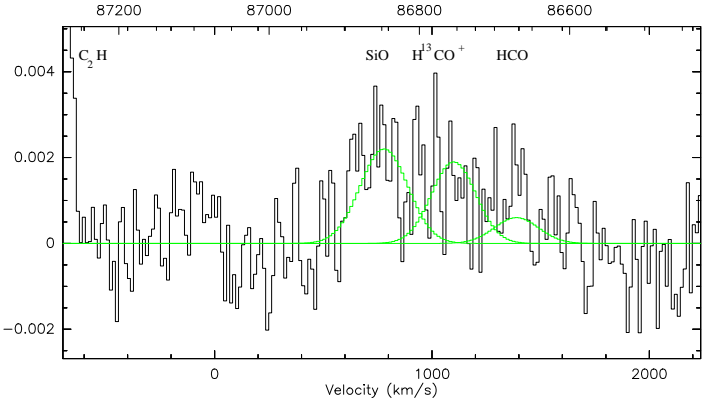}
\caption{Gaussian fits to the blended lines of H$^{13}$CO$^+$\,$(1-0)$, HCO\,$(1_{1,0}-0_{1,0})$ and SiO\,$(2-1)$. We first fitted a Gaussian profile to the H$^{13}$CO$^+$ line, and then removed its emission. After that, we fitted the hyperfine structure of HCO, and also removed it. Finally, we fitted a SiO Gaussian profile to the residuals. See more details in the text.}
\label{SiO}
\end{center}
\end{figure}

\begin{figure}[t!]
\begin{center}
\includegraphics[angle=0,width=0.4\textwidth]{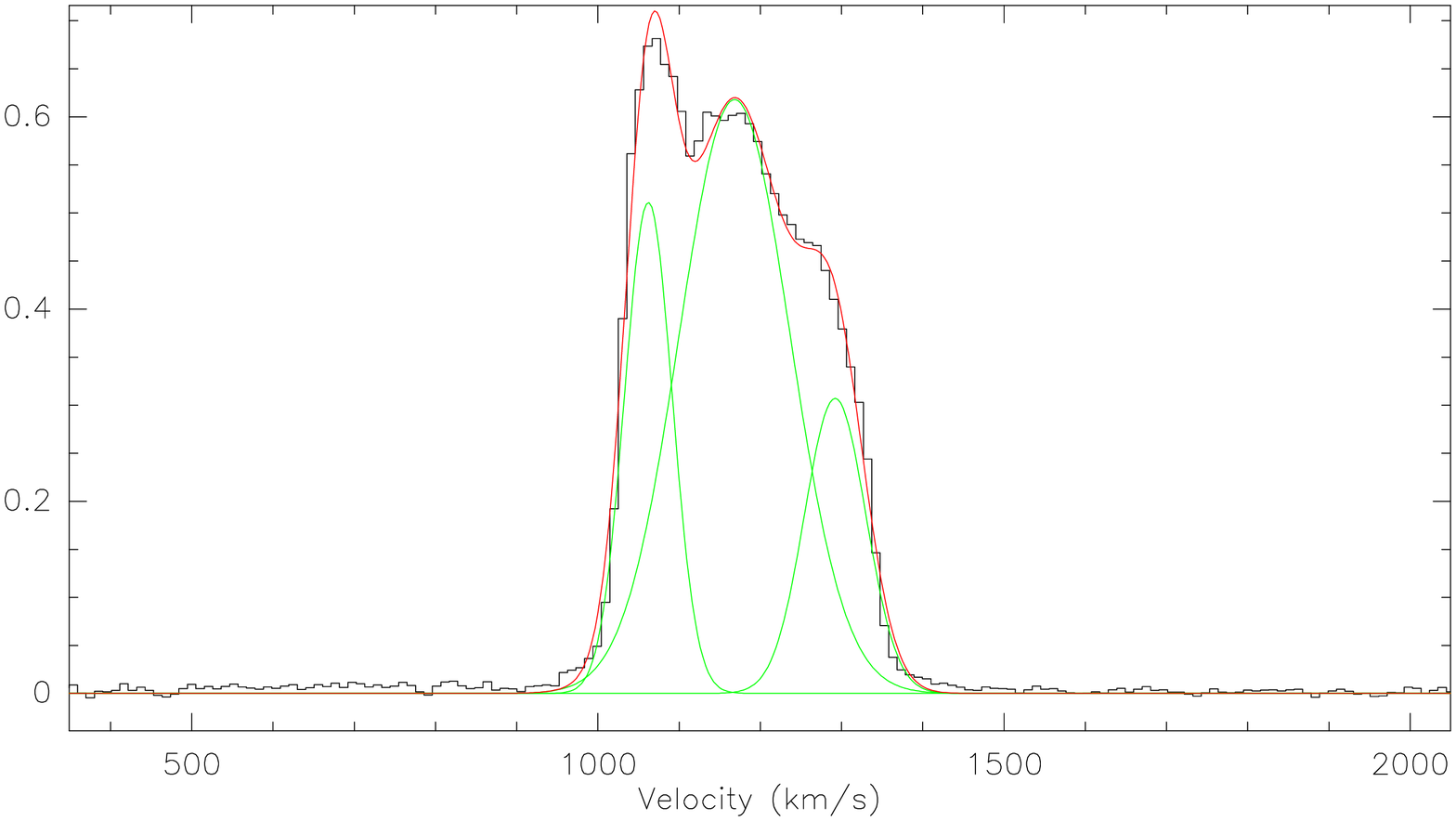}
\caption{Line profile of the CO\,$(1-0)$ line. Y-xis is in $T_{\rm mb}$ scale. The contribution of the starburst ring emission was fitted by two Gaussians at the wings of the central Gaussian, which contains the CND emission.}
\label{CO}
\end{center}
\end{figure}

\item{\bf Methyl cyanide - CH$_3$CN}\\
This symmetric top molecule is detected for the first time in NGC\,1068.  The two CH$_3$CN\,$(5_k-4_k)$ and CH$_3$CN\,$(6_k-5_k)$ transitions at 92.0\,GHz and 110.4\,GHz lie inside our frequency coverage. Each of them are formed by 25 spectral features with the quantum number $k$ going from 0 to $J-1$. Although faint, both lines are strong enough to be detected with a S/N\,$\geq$\,3. The transition at 110.4\,GHz is blended with HNCO\,$(5_{1,4}-4_{1,3})$. We fitted a synthetic Gaussian profile to both lines, fixing the line position of CH$_3$CN to 1100\,km\,s$^{-1}$ and the linewidth to 150\,km\,s$^{-1}$ (based on the non-blended line). Its rotational temperature is similar in NGC\,1068 ($T_{\rm rot}\sim29$\,K) and M\,82 ($T_{\rm rot}\sim33$\,K).

\item{\bf Sulphur monoxide - SO}\\
Three transitions of sulphur monoxide were detected for the first time in NGC\,1068 at 99.3\,GHz, 100.0\,GHz, and 109.252\,GHz. The first one was not blended and its Gaussian parameters have been used to fit the other two. The line is narrow, 136\,km\,s$^{-1}$. The other two lines are blended with HC$_3$N$(11-10)$ and HC$_3$N$(12-11)$ repectively. The contribution of SO was firstly subtracted and then we fitted cyanoacetylene to the residuals. The SO\,$(5_4-4_4)$ at 100.0\,GHz does not fit well in the rotational diagram, so the values we show in Table~\ref{TableNT} do not take into account this transition. 

\item{\bf Hydroxymethylidynium - HOC$^+$}\\
We detected the HOC$^+(1-0)$ transition at 89.5\,GHz.  Its peak intensity temperature is slightly above the noise, so we consider it as a tentative detection. This line shows a quite narrow linewidth, of about 128\,km\,s$^{-1}$. Fixing the rotational temperature to $10\pm5$\,K, we obtained the lowest column density of the molecular survey, with $N_{\rm HOC^+}$\,=\,(1.1$\pm$0.6)$\times10^{13}$\,cm$^{-2}$.

\item{\bf Cyanoacetylene - HC$_3$N}\\
This molecule is detected for the first time in NGC\,1068. We identified  HC$_3$N\,$(10-9)$ at 91.0\,GHz, HC$_3$N\,$(11-10)$ at 100.1\,GHz, and HC$_3$N\,$(12-11)$ at 109.2\,GHz. The transition at 91.0\,GHz was not blended, so we used its Gaussian parameters to fix the other two, which were blended with SO. For those lines, we firstly subtracted the sulphur monoxide features (as explained before), and then fitted two synthetic Gaussian profiles to the residuals. The HC$_3$N rotational temperature in NGC\,1068 ($7.3\pm1.4$\,K) is significantly lower than the ones found for M\,82 and NGC\,253 ($\sim$25\,K and $\sim$[12-73]\,K respectively, \citealt{Aladro10}).

\item{\bf Diazenylium - N$_2$H$^+$}\\
This is the first time that this species is detected in NGC\,1068. Only the N$_2$H$^+(1-0)$ transition, at 93.2\,GHz, lies inside our survey. It is formed by three features, so we performed a synthetic Gaussian fit in order to calculate the contribution of each one. We fixed the rotational temperature to $10\pm5$\,K to calculate its column density.

\item{\bf Carbon monosulfide - CS and C$^{34}$S}\\
We detected the CS\,$(2-1)$ line, which is blended with $^{13}$CO\,$(1-0)$  coming from the image band. Since these two are the only transitions of these  molecules detected in our survey, it is not possible to dissentangle the contribution of each one. Thus, no Gaussian fit was done. Nevertheless, we used the CS\,$(3-2)$, CS\,$(5-4)$, and CS\,$(7-6)$ lines observed by \citet{Bayet09b} to do the rotation diagram. Two gas components with different temperatures and column densities are seen.

C$^{34}$S\,$(2-1)$ was detected at 96.4\,GHz. This transition was complemented with the C$^{34}$S\,$(3-2)$ line observed by \citet{Martin09a}. The rotational temperature is one of the lowest obtained in the survey, with only $\sim4$\,K.

\item{\bf Carbon monoxide - CO, C$^{18}$O, and $^{13}$CO}\\
CO\,$(1-0)$, at 115.3\,GHz, was blended with NS\,$(3_1-2_1)$, but since NS was comparatively very faint, we assumed its contribution negligible. As the emission of CO\,$(1-0)$ is quite spread in the nucleus of NGC\,1068, it clearly shows a deviation from a Gaussian profile, which points to the complex gas dynamics in that region, as well as a possible contribution of the starburst ring emission. We fitted a three Gaussian profile, which is shown in Fig.~\ref{CO}. 

We also used the CO\,$(2-1)$, CO\,$(3-2)$, and CO\,$(4-3)$ lines from \citet{Israel09}. Like CS, CO shows two components in the rotation diagram. Despite CO\,$(1-0)$ is the strongest feature in the survey, it is not seen in the image band. This is due to the variable rejection along the 3\,mm band.

$^{13}$CO\,$(1-0)$ was detected at 110.2\,GHz. Similarly to $^{12}$CO\,$(1-0)$, this transition showed a deviation from a Gaussian profile. It was complemented with the $^{13}$CO\,$(2-1)$ and $^{13}$CO\,$(3-2)$ lines observed by \citet{Israel09}. Two different components are also seen in the Boltzmann diagram. As commented before, $^{13}$CO \,$(1-0)$ does appear in the image band, blended with CS\,$(2-1)$.

C$^{18}$O\,$(1-0)$ appears at 109.8\,GHz. It was not blended, but it showed a bump at the higher frequencies due to the proximity of a HNCO transition. To avoid the isocyanic acid contamination, we fixed the linewidth to 230\,km\,s$^{-1}$ and then fitted three Gaussians, which better account for the line profile. We also used the C$^{18}$O\,$(2-1)$ line observed by \citet{Martin09a} for complementing our observations. The rotational temperature is the lowest one among all the molecules  detected ($T_{\rm rot}=3.3\pm0.1$\,K).

\item{\bf Cyanogen - CN}\\
Two separated lines of cyanogen were identified at 113.2\,GHz (CN\,$(1_{0,1}-0_{0,1})$) and 113.5\,GHz (CN\,$(1_{0,2}-0_{0,1})$). The first one is formed by four different features, while the second one has five. We fitted a synthetic Gaussian profile to both lines. We included other CN transitions observed by \citet{Perez09}. We also detected two multi-feature lines of  $^{13}$CN at 108.4\,GHz and 108.6\,GHz.

\item{\bf Methanol - CH$_3$OH}\\
Only the  CH$_3$OH\,$(2_k-1_k)$ line is seen at 96.7\,GHz, which is composed by four features with the quantum number $k$ varying between $-1,0,1$. However, we calculated the integrated intensities of other transitions that lie inside our frequency range, which fulfil the conditions imposed to the energy and logarithm of the integrated intensities, as explained at the beginning of this appendix. Taking into account all these features (thirteen in total), we performed a synthetic Gaussian fit. 

\item{\bf Nitrogen monosulfide - NS}\\
We detected the NS\,$(3_1-2_1)$ transition in our survey. It is splitted in two separated multi-lines with six features each. The first one is blended with CO\,$(1-0)$, while the second one is located at higher frequencies, so that can be easily separated from CO. We firstly subtracted the carbon monoxide contribution, and then fitted a synthetic Gaussian profile to all the NS features. We assumed $T_{\rm rot}=10\pm5$\,K to calculate the column density.

\end{itemize}

\clearpage

\Online

\onecolumn
\begin{figure*}[!ht]
\begin{center}
\includegraphics[angle=0,width=\textwidth]{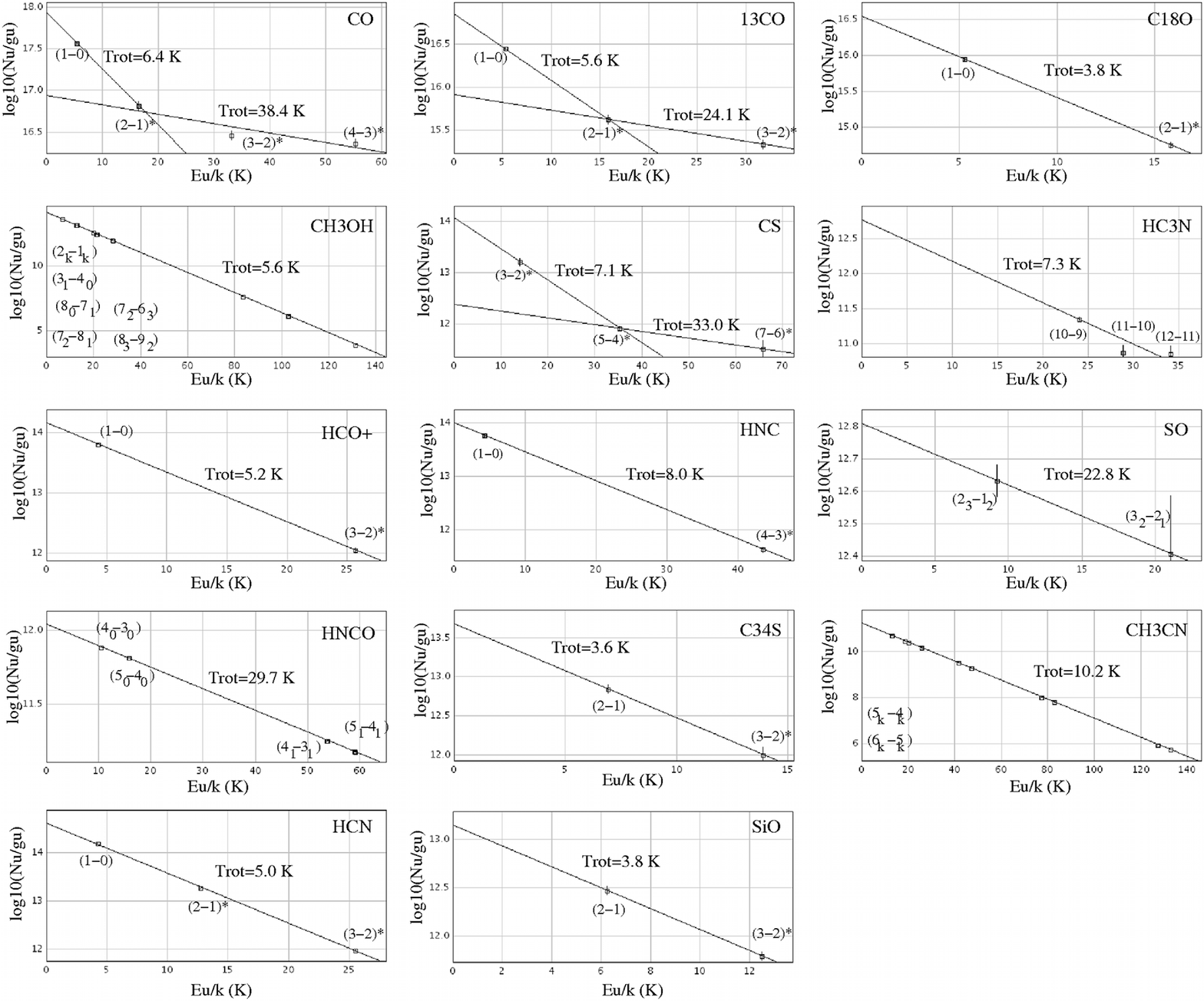}

\caption{Boltzmann diagrams of the molecules observed in the survey for which we have detected more than one transition, or for which we have taken other transitions from the literature (marked with *). The resulting rotational temperatures are indicated.}

\label{DR}
\end{center}
\end{figure*}
\end{appendix}


\clearpage

\onecolumn
\begin{appendix}
\section{Gaussian fits parameter results.}

\begin{table*}[h!]
\centering
\begin{tabular}[!h]{lccccccccccc} 
\hline
\hline
Line	&	Frequency	&	$I$ &    Position 	  &      Width 		& Tpeak		& Comments	\\
	&	MHz		&	K\,km\,s$^{-1}$		&  km\,s$^{-1}$	  &   km\,s$^{-1}$	& mK		&		\\	
\hline
H$^{13}$CN\,$(1-0)$ &	86340.2	& 	$0.4\pm...$	&   $1066.3\pm0.0$	&  $166.4\pm0.0$	& 2.4  		& hf, s 	\\ 

HCO\,$(1_{1,0}-0_{1,0})$&	86670.8		& $0.2\pm...$	& $1100.0\pm0.0$		& $240.0\pm0.0$		& 0.6		& b, s\\
H$^{13}$CO$^+$\,$(1-0)$&	86754.3		& $0.5\pm...$	&  $1100.0\pm0.0$	& $240.0\pm0.0$	& 1.9		& b, s \\
SiO\,$(2-1)$	&	86847.0		& 	$0.6\pm...$	&  $1100.0\pm0.0$	& $259.1\pm0.0$	& 2.2		& b, s \\
HN$^{13}$C\,$(1-0)$ &	87090.8		& 	$0.2\pm0.3$	&  $1075.7\pm124.4$  	& $151.8\pm235.9$	&  1.1 		&  \\ 
C$_2$H\,$(1-0)$	&	87316.9		& 	$3.1\pm...$	&  $1098.4\pm0.0$	& $240.0\pm0.0$		&  12.0		& hf, s \\ 
HNCO\,$(4_{0,4}-3_{0,3})$	&	87925.2		& 	$0.1\pm...$	&  $1100.0\pm0.0$	& $230.0\pm0.0$		& 0.4		& m, s	\\
HCN\,$(1-0)$	&	88631.8		& 	$11.1\pm...$	&  $1120.1\pm0.0$	& $238.5\pm0.0$		&  43.0		& hf, s	\\ 
HCO$^+$\,$(1-0)$ &	89188.6		& 	$13.8\pm0.3$ 	&  $1122.6\pm1.2$ 	& $230.9\pm1.5$		&  55.9  	&	\\ 
HOC$^+(1-0)$	&	89487.4		&	$0.2\pm 0.1$  	&  $1173.3 \pm17.5$	& $128.2\pm51.6$	&  1.8		&	\\%
HNC$(1-0)$	&	90663.6		& 	$7.5\pm0.6$	&  $1141.3\pm0.3$ 	& $235.9\pm3.2$		& 30.8		&	\\
HC$_3$N\,$(10-9)$	&  	90979.0		&	 $1.1\pm0.1$	&  $1142.6\pm10.0$	& $263.7\pm24.9$	& 3.8		&	\\

CH$_3$CN$\,(5_k-4_k)$&	91987.0		& 	$0.1\pm...$	& $1100.0\pm0.0$	& $150.0\pm0.0$		& 0.6		& m, s	\\ 

N$_2$H$^+$\,$(1-0)$ &	93173.7		& 	$1.0\pm...$	& $1164.1\pm0.0$	& $233.5\pm0.0$		& 4.0		& m, s  \\
C$^{34}$S\,$(2-1)$ &	96412.9		& 	$0.7\pm0.1$	& $1091.2\pm15.7$ 	& $249.0\pm29.5$	& 2.7		& 	\\
CH$_3$OH\,$(2_k-1_k)$&	96741.4		&	$0.9\pm...$	& $1117.5\pm0.0$	& $247.2\pm0.0$		& 3.3		& m, s	\\
CS\,$(2-1)$	&	97981.0		& ...			& ...			& ...			& ...		& b	\\
SO\,$(2_1-3_2)$	&	99299.9		& 	$0.4\pm0.1$	&  $1111.1\pm7.5$	& $136.0\pm17.0$	&  3.1		&	\\
SO\,$(5_4-4_4)$	&	100029.6 	& 	$0.2\pm0.1$	&  $1111.1\pm0.0$	& $136.0\pm0.0$		& 1.6		& b\\ 
HC$_3$N\,$(11-10)$	&  100076.4	& 	$0.5\pm...$	&  $1142.6\pm0.0$	& $263.0\pm0.0$		& 2.0		& b, s\\ 
$^{13}$CN\,$(1_{1,0}-0_{1,1})$ & 108426.9	&	$0.1\pm...$	&  $1104.\pm0.0$	& $257.2\pm0.0$		& 0.2		& m, s \\	
$^{13}$CN\,$(1_{2,1}-0_{1,0})$ & 108651.3	&	$0.2\pm...$	&  $1104.\pm0.0$	& $257.2\pm0.0$		& 0.6		& m, s \\	
HC$_3$N\,$(12-11)$	&  	109173.6	& 	$0.7\pm...$	& $1142.6\pm0.0$	& $263.0\pm0.0$		& 2.6		& b, s\\ 
SO\,$(3_2-2_1)$	&	109252.2	& 	$0.2\pm0.1$	& $1111.1\pm0.0$	& $136.0\pm0.0$		& 1.3		& b\\ 
HNCO\,$(5_{1,5}-4_{1,4})$	&	109496.0	& 	$0.04\pm...$	& $1100.0\pm0.0$	& $230.0\pm0.0$ 	& 0.2		& m, s	\\
C$^{18}$O\,$(1-0)$&	109782.2	& 	$3.6\pm0.2$ 	& $1142.2\pm6.1$	& $230.0\pm0.0$		& 13.0		&	\\ 
HNCO\,$(5_{0,5}-4_{0,4})$	&	109905.8	& 	$0.3\pm...$	& $1100.0\pm0.0$	& $230.0\pm0.0$		& 1.3		& m, s \\ 
$^{13}$CO\,$(1-0)$&	110201.4	& 	$12.8\pm0.7$	& $1105.2\pm3.6$	& $252.1\pm4.9$ 	& 45.8		&  \\ 
HNCO\,$(5_{1,4}-4_{1,3})$	&	110298.1	& 	$0.2\pm...$	& $1100.0\pm0.0$	& $230.0\pm0.0$		& 0.7		& m, b, s	\\
CH$_3$CN\,$(6_k-5_k)$&	110383.5	& 	$0.1\pm...$	& $1100.0\pm0.0$	& $150.0\pm0.0$		& 0.8		& m, b, s\\ 
CN\,$(1_{0,2}-0_{0,1})$	&113191.3	& 	$4.5\pm...$	& $1152.8\pm0.0$	& $247.4\pm0.0$	 	& 17.3		& s\\ 
CN\,$(1_{0,2}-0_{0,1})$	&113491.0	& 	$12.1\pm...$	& $1152.8\pm0.0$	& $247.4\pm0.0$		& 46.0		& s\\ 
NS\,$(3_1-2_1)$			& 115153.9	&	$0.4\pm...$	& $1100.0\pm0.0$	& $140.0\pm0.0$		& 2.8		& m, b, s \\
$^{12}$CO\,$(1-0)$ &	115271.2	& 	$172.2\pm4.5$	& $1168.3\pm0.3$	& $251.5\pm0.4$		& 618.0		& b	\\ 
NS\,\,$(3_1-2_1)$	& 115556.2	& 	$0.4\pm...$	& $1100.0\pm0.0$	& $140.0\pm0.0$		& 2.8		& m, s \\	
\hline
\end{tabular}

\caption{Parameters fixed during the Gaussian fit have zero errors associated. Synthetic fits do not give errors to the integrated area. Remarks: $(b)$ blended line; $(m)$ multi-transition line; $(hf)$ hyperfine transition; $(s)$ synthetic Gaussian fit using MASSA. For those transitions showing several components, the parameters refer to the main component of the group.}
\label{TableGauss}
\end{table*}

\end{appendix}


\begin{thebibliography}{}


\bibitem[Aladro et al.(2011a)]{Aladro10} Aladro, R., Mart{\'{\i}}n-Pintado, J., Mart{\'{\i}}n, S., Mauersberger, R., \& Bayet, E.\ 2011a, \aap, 525, A89 

\bibitem[Aladro et al.(2011b)]{Aladro11} Aladro, R., Mart{\'{\i}}n, S., Mart{\'{\i}}n-Pintado, J., et al.\ 2011b, \aap, 535, A84 

\bibitem[Ao et al.(2011)]{Ao2011} Ao, Y., Henkel, C., Braatz, J.~A., et al.\ 2011, \aap, 529, A154

\bibitem[Bayet et al.(2008)]{Bayet08} Bayet, E., Viti, S., Williams, D.~A., \& Rawlings, J.~M.~C.\ 2008, \apj, 676, 978 


\bibitem[Bayet et al.(2009a)]{Bayet09a} Bayet, E., Viti, S., Williams, D.~A., Rawlings, J.~M.~C., \& Bell, T.\ 2009a, \apj, 696, 1466 

\bibitem[Bayet et al.(2009b)]{Bayet09b} Bayet, E., Aladro, R., Mart{\'{\i}}n, S., Viti, S., \& Mart{\'{\i}}n-Pintado, J.\ 2009b, \apj, 707, 126 

\bibitem[Blake et al.(1987)]{Blake87} Blake, G.~A., Sutton, E.~C., Masson, C.~R., \& Phillips, T.~G.\ 1987, \apj, 315, 621 

\bibitem[Bland-Hawthorn et al.(1997)]{Bland-Hawthorn97} Bland-Hawthorn, J., Gallimore, J.~F., Tacconi, L.~J., Brinks, E., Baum, S.~A., Antonucci, R.~R.~J., \& Cecil, G.~N.\ 1997, \apss, 248, 9   

\bibitem[Chin et al.(1996)]{Chin96} Chin, Y.-N., Henkel, C., Whiteoak, J.~B., Langer, N., \& Churchwell, E.~B.\ 1996, \aap, 305, 960 

\bibitem[Chu \& Watson(1983)]{Chu83} Chu, Y.-H., \& Watson, W.~D.\ 1983, \apj, 267, 151 

\bibitem[Dalgarno(2006)]{Dalgarno06} Dalgarno, A.\ 2006, 
Proceedings of the National Academy of Science, 103, 12269 

\bibitem[Davies et al.(2007)]{Davies07} Davies, R.~I., S{\'a}nchez, F.~M., Genzel, R., Tacconi, L.~J., Hicks, E.~K.~S., Friedrich, S., \& Sternberg, A.\ 2007, \apj, 671, 1388   

\bibitem[Fern{\'a}ndez-Ontiveros et al.(2009)]{Onti09} Fern{\'a}ndez-Ontiveros, J.~A., Prieto, M.~A., \& Acosta-Pulido, J.~A.\ 2009, \mnras, 392, L16 

\bibitem[Fontani et al.(2007)]{Fontani07} Fontani, F., Pascucci, I., Caselli, P., et al.\ 2007, \aap, 470, 639 

\bibitem[Freedman et al.(1994)]{Freedman94} Freedman, W.~L., et al.\ 1994, \apj, 427, 628 

\bibitem[Frerking et al.(1980)]{Frerking80} Frerking, M.~A., Wilson, R.~W., Linke, R.~A., \& Wannier, P.~G.\ 1980, \apj, 240, 65 

\bibitem[Garc{\'{\i}}a-Burillo et al.(2010)]{Burillo10} Garc{\'{\i}}a-Burillo, S., et al.\ 2010, \aap, 519, A2 

\bibitem[Glassgold et al.(1985)]{Glassgold85} Glassgold, A.~E., Huggins, P.~J., \& Langer, W.~D.\ 1985, \apj, 290, 615 

\bibitem[Goldsmith \& Langer(1999)]{Goldsmith99} Goldsmith, P.~F., \& Langer, W.~D.\ 1999, \apj, 517, 209 

\bibitem[Hatchell et al.(1998)]{Hatchell98} Hatchell, J., Thompson, M.~A., Millar, T.~J., \& MacDonald, G.~H.\ 1998, \aaps, 133, 29 

\bibitem[Helfer \& Blitz(1995)]{Helfer95} Helfer, T.~T., \& Blitz, L.\ 1995, \apj, 450, 90 

\bibitem[Henkel \& Mauersberger(1993)]{HenkelMauers93} Henkel, C., \& Mauersberger, R.\ 1993, \aap, 274, 730  

\bibitem[Huggins et al.(1984)]{Huggins84b} Huggins, P.~J., Glassgold, A.~E., \& Morris, M.\ 1984, \apj, 279, 284 

\bibitem[Israel(2009)]{Israel09} Israel, F.~P.\ 2009, \aap, 493, 525 

\bibitem[Jackson et al.(1993)]{Jackson93} Jackson, J.~M., Paglione, T.~A.~D., Ishizuki, S., \& Nguyen-Q-Rieu 1993, \apjl, 418, L13 

\bibitem[Jaffe et al.(2004)]{Jaffe04} Jaffe, W., Meisenheimer, K., R{\"o}ttgering, H.~J.~A., et al.\ 2004, \nat, 429, 47 

\bibitem[Jim{\'e}nez-Serra et al.(2008)]{Izaskun09} Jim{\'e}nez-Serra, I., Caselli, P., Mart{\'{\i}}n-Pintado, J., \& Hartquist, T.~W.\ 2008, \aap, 482, 549 

\bibitem[Kamenetzky et al.(2011)]{Kamenetzky11} Kamenetzky, J., Glenn, J., Maloney, P.~R., et al.\ 2011, \apj, 731, 83 

\bibitem[Karachentsev et al.(2003)]{Karachentsev03} Karachentsev, I.~D., Grebel, E.~K., Sharina, M.~E., et al.\ 2003, \aap, 404, 93 

\bibitem[Kohno et al.(2001)]{Kohno01} Kohno, K., Matsushita, S., Vila-Vilar{\'o}, B., et al.\ 2001, The Central Kiloparsec of Starbursts and AGN: The La Palma Connection, 249, 672 

\bibitem[Konstantopoulos et al.(2009)]{Konstantopoulos09} Konstantopoulos, I.~S., Bastian, N., Smith, L.~J., Westmoquette, M.~S., Trancho, G., \& Gallagher, J.~S.\ 2009, \apj, 701, 1015 

\bibitem[Krips et al.(2008)]{Krips08} Krips, M., Neri, R., Garc{\'{\i}}a-Burillo, S., et al.\ 2008, \apj, 677, 262 

\bibitem[Krips et al.(2011)]{Krips11} Krips, M., Mart{\'{\i}}n, S., Eckart, A., et al.\ 2011, \apj, 736, 37 

\bibitem[Le Teuff et al.(2000)]{LeTeuff00} Le Teuff, Y.~H., Millar, T.~J., \& Markwick, A.~J.\ 2000, \aaps, 146, 157 

\bibitem[Lovas(1992)]{Lovas92} Lovas, F.~J.\ 1992, J. Phys. Chem. Ref. Data, 21, 181 

\bibitem[Maloney et al.(1996)]{Maloney96} Maloney, P.~R., Hollenbach, D.~J., \& Tielens, A.~G.~G.~M.\ 1996, \apj, 466, 561 

\bibitem[Mart{\'{\i}}n et al.(2006)]{Martin06b} Mart{\'{\i}}n, S., Mart{\'{\i}}n-Pintado, J., \& Mauersberger, R.\ 2006, A\&A, 450, L13      

\bibitem[Mart{\'{\i}}n et al.(2009)]{Martin09a} Mart{\'{\i}}n, S., Mart{\'{\i}}n-Pintado, J., \& Mauersberger, R.\ 2009, \apj, 694, 610 

\bibitem[Mart{\'{\i}}n et al.(2010)]{Martin10} Mart{\'{\i}}n, S., Aladro, R., Mart{\'{\i}}n-Pintado, J., \& Mauersberger, R.\ 2010, \aap, 522, A62 

\bibitem[Meijerink \& Spaans(2005)]{Meijerink05} Meijerink, R., \& Spaans, M.\ 2005, \aap, 436, 397 

\bibitem[Meijerink et al.(2011)]{Meijerink11} Meijerink, R., Spaans, M., Loenen, A.~F., \& van der Werf, P.~P.\ 2011, \aap, 525, A119 

\bibitem[Millar et al.(1997)]{Millar97} Millar, T.~J., MacDonald, G.~H., \& Gibb, A.~G.\ 1997, \aap, 325, 1163 

\bibitem[Millar \& Hatchell(1998)]{Millar98} Millar, T.~J., \& Hatchell, J.\ 1998, Faraday Discussions, 109, 15 

\bibitem[M{\"u}hle et al.(2007)]{Muhle07} M{\"u}hle, S., Seaquist, E.~R., \& Henkel, C.\ 2007, \apj, 671, 1579 

\bibitem[M{\"u}ller et al.(2001)]{Muller01} M{\"u}ller, H.~S.~P., Thorwirth, S., Roth, D.~A., \& Winnewisser, G.\ 2001, \aap, 370, L49 

\bibitem[M{\"u}ller et al.(2005)]{Muller05} M{\"u}ller, H.~S.~P., Schl{\"o}der, F., Stutzki, J., \& Winnewisser, G.\ 2005, Journal of Molecular Structure, 742, 215 

\bibitem[M{\"u}ller S{\'a}nchez et al.(2009)]{MullerSanchez09} M{\"u}ller S{\'a}nchez, F., Davies, R.~I., Genzel, R., et al.\ 2009, \apj, 691, 749 

\bibitem[Nakajima et al.(2011)]{Nakajima11} Nakajima, T., Takano, S., Kohno, K., \& Inoue, H.\ 2011, \apjl, 728, L38 

\bibitem[Papadopoulos et al.(1996)]{Papadopoulos96} Papadopoulos, P.~P., Seaquist, E.~R., \& Scoville, N.~Z.\ 1996, \apj, 465, 173 

\bibitem[Papadopoulos \& Seaquist(1999)]{Papadopoulos99} Papadopoulos, P.~P., \& Seaquist, E.~R.\ 1999, \apj, 516, 114 

\bibitem[Penzias \& Burrus(1973)]{Penzias73} Penzias, A.~A., \& Burrus, C.~A.\ 1973, \araa, 11, 51 

\bibitem[P{\'e}rez-Beaupuits et al.(2009)]{Perez09} P{\'e}rez-Beaupuits, J.~P., Spaans, M., van der Tak, F.~F.~S., et al.\ 2009, \aap, 503, 459 

\bibitem[Pickett et al.(1998)]{Pickett88} Pickett, H.~M., Poynter, I.~R.~L., Cohen, E.~A., et al.\ 1998, J. of Quantitative Spect. and Rad. Transf., 60, 883   

\bibitem[Sakai et al.(2010)]{Sakai10} Sakai, N., Saruwatari, O., Sakai, T., Takano, S., \& Yamamoto, S.\ 2010, \aap, 512, A31 

\bibitem[Schinnerer et al.(2000)]{Schinnerer00} Schinnerer, E., Eckart, A., Tacconi, L.~J., Genzel, R., \& Downes, D.\ 2000, \apj, 533, 850 

\bibitem[Spaans \& Meijerink(2005)]{Spaans05} Spaans, M., \& Meijerink, R.\ 2005, \apss, 295, 239 

\bibitem[Sternberg \& Dalgarno(1995)]{Sternberg95} Sternberg, A., \& Dalgarno, A.\ 1995, \apjs, 99, 565 

\bibitem[Sternberg et al.(1994)]{Sternberg94} Sternberg, A., Genzel, R., \& Tacconi, L.\ 1994, \apjl, 436, L131 

\bibitem[Tacconi et al.(1994)]{Tacconi94} Tacconi, L.~J., Genzel, R., Blietz, M., et al.\ 1994, \apjl, 426, L77 

\bibitem[Telesco \& Harper(1980)]{Telesco80} Telesco, C.~M., \& Harper, D.~A.\ 1980, \apj, 235, 392 

\bibitem[Usero et al.(2004)]{Usero04} Usero, A., Garc{\'{\i}}a-Burillo, S., Fuente, A., Mart{\'{\i}}n-Pintado, J., \& Rodr{\'{\i}}guez-Fern{\'a}ndez, N.~J.\ 2004, \aap, 419, 897 

\bibitem[Viti \& Williams (1999)]{Viti99} Viti, S., \& Williams, D.~A.\ 1999, MNRAS, 305, 755 

\bibitem[Viti et al.(2004)]{Viti04} Viti, S., Collings, M.~P., Dever, J.~W., McCoustra, M.~R.~S., \& Williams, D.~A.\ 2004, MNRAS, 354, 1141 

\bibitem[Viti et al.(2011)]{Viti11} Viti, S., Jimenez-Serra, I., Yates, J.~A., et al.\ 2011, \apjl, 740, L3 

\bibitem[Wannier(1980)]{Wannier80} Wannier, P.~G.\ 1980, \araa, 18, 399 

\bibitem[Wilson \& Rood(1994)]{Wilson94} Wilson, T.~L., \& Rood, R.\ 1994, \araa, 32, 191 

\bibitem[Woodall et al.(2007)]{Woodall07} Woodall, J., Ag{\'u}ndez, M., Markwick-Kemper, A.~J., \& Millar, T.~J.\ 2007, A\&A, 466, 1197 

\bibitem[Woods \& Willacy(2009)]{Woods09} Woods, P.~M., \& Willacy, K.\ 2009, \apj, 693, 1360 

\bibitem[Young \& Sanders(1986)]{Young86} Young, J.~S., \& Sanders, D.~B.\ 1986, \apj, 302, 680 

\bibitem[Zaritsky et al.(1994)]{Zaritsky94} Zaritsky, D., Kennicutt, R.~C., Jr., \& Huchra, J.~P.\ 1994, ApJ, 420, 87 

\end{thebibliography}
\end{document}